\documentclass[fleqn, usenatbib]{aa}

\usepackage[T1]{fontenc}
\usepackage{txfonts}
\usepackage[switch]{lineno}

\DeclareRobustCommand{\VAN}[3]{#2}
\let\VANthebibliography\thebibliography
\def\thebibliography{\DeclareRobustCommand{\VAN}[3]{##3}\VANthebibliography}

\usepackage{graphicx}   
\usepackage{amsmath}    
\usepackage{amssymb}    
\usepackage{newtxtext,newtxmath}
\usepackage{tabularx}
\usepackage{xspace}
\usepackage{multicol, multirow}
\usepackage{lipsum}
\usepackage{hyperref}

\newcommand{\cigale}{\textsc{cigale}\xspace}
\newcommand{\prospector}{\textsc{Prospector}\xspace}
\newcommand{\palpha}{\textsc{Prospector}-$\alpha$\xspace}
\newcommand{\fsps}{\textsc{FSPS}\xspace}

\newcommand{\agemw}{$\langle t_* \rangle_m$\xspace}

\hypersetup{colorlinks, citecolor=blue,
            linkcolor=blue, urlcolor=blue}

\begin{document}

\title{The PAU Survey: Galaxy stellar population properties estimates with narrowband data}

   \author{B. Csizi\inst{1,2}
          \thanks{e-mail: benjamin.csizi@uibk.ac.at}
          \and
          L. Tortorelli\inst{2,3}
          \and
          M. Siudek\inst{4,5}
          \and 
          D. Grün\inst{2,3}
          \and 
          P. Renard\inst{6}
          \and
          P. Tallada-Crespí\inst{7,8}
          \and
          E. Sánchez\inst{7}
          \and
          R. Miquel\inst{9,10}
          \and
          C. Padilla\inst{10}
          \and
          J. García-Bellido\inst{11}
          \and
          E. Gaztañaga\inst{12,13,4}
          \and
          R. Casas\inst{13,4}
          \and
          S. Serrano\inst{14,4,13}
          \and
          J. De Vicente\inst{7}
          \and
          E. Fernandez\inst{10}
          \and
          M. Eriksen\inst{10,8}
          \and
          G. Manzoni\inst{15}
          \and
          C. M. Baugh\inst{16}
          \and
          J. Carretero\inst{7,8}
          \and
          F. J. Castander\inst{13,4}
          }

   \institute{Universit\"at Innsbruck, Institut f\"ur Astro- und Teilchenphysik, Technikerstr. 25/8, 6020 Innsbruck, Austria
        \and
         University Observatory, Faculty of Physics, Ludwig-Maximilians-Universität, Scheinerstr. 1, 81677 Munich, Germany
        \and 
         Excellence Cluster ORIGINS, Boltzmannstr. 2, 85748 Garching, Germany
        \and
         Institute of Space Sciences (ICE, CSIC), Campus UAB, Carrerde Can Magrans, 08193 Bellaterra (Barcelona), Spain
        \and
        Instituto Astrofísica de Canarias, Av. Via Láctea s/n, E38205 La Laguna, Spain
        \and 
        Department of Astronomy, Tsinghua University, Beijing 100084, China
        \and
         Centro de Investigaciones Energéticas, Medioambientales y Tecnológicas (CIEMAT), Avenida Complutense 40, E-28040 Madrid (Madrid), Spain
        \and
        Port d'Informació Científica (PIC), Campus UAB, C. Albareda s/n, 08193 Bellaterra (Cerdanyola del Vallès), Spain
        \and
        Institució Catalana de Recerca i Estudis Avançats (ICREA), Pg. de Lluís Companys 23, 08010 Barcelona, Spain 
        \and
        Institut de Física d'Altes Energies (IFAE), The Barcelona Institute of Science and Technology, Campus UAB, 08193 Bellaterra (Barcelona) Spain 
        \and 
         Instituto de Fisica Teorica (UAM/CSIC), Universidad Autonoma de Madrid, E-28049 Madrid, Spain
         \and
         Institute of Cosmology and Gravitation (University of Portsmouth), Portsmouth, UK
         \and
        Institut d'Estudis Espacials de Catalunya (IEEC), Gran Capità 2-4, Ed. Nexus 201, 08034 Barcelona, Spain
         \and
         Satlantis, University Science Park, Sede Bld 48940, Leioa-Bilbao, Spain
         \and
         Jockey Club Institute for Advanced Study, The Hong Kong University of Science and Technology, Hong Kong S.A.R., China
         \and
         Institute for Computational Cosmology, Department of Physics, Durham University, South Road, Durham DH1 3LE, UK
    }

   \date{Received xx, 2024}

  \abstract{
  A newfound interest has been seen in narrowband galaxy surveys as a promising method for achieving the necessary accuracy on the photometric redshift estimate of individual galaxies for next-generation stage IV cosmological surveys. One key advantage is the ability to provide higher spectral resolution information on galaxies, which ought to allow for a more accurate and precise estimation of the stellar population properties for galaxies. However, the impact of adding narrowband photometry on the stellar population properties estimate is largely unexplored. The scope of this work is two-fold: 1) we leverage the predictive power of broadband and narrowband data to infer galaxy physical properties, such as stellar masses, ages, star formation rates, and metallicities; and 2) we evaluate the improvement of performance in estimating galaxy properties when we use narrowband  instead of broadband data. In this work, we measured the stellar population properties of a sample of galaxies in the COSMOS field for which both narrowband and broadband data are available. In particular, we employed narrowband data from the Physics of the Accelerating Universe Survey (PAUS) and broadband data from the Canada France Hawaii Telescope legacy survey (CFHTLS). We used two different spectral energy distribution (SED) fitting codes to measure galaxy properties, namely, \cigale and \prospector. We find that the increased spectral resolution of narrowband photometry does not yield a substantial improvement in terms of constraining the galaxy properties using the SED fitting. Nonetheless, we find that we are able to obtain a more diverse distribution of metallicities and dust optical depths with \cigale when employing the narrowband data. The effect is not as prominent as expected, which we relate to the low narrowband signal-to-noise ratio (S/N) of a majority of the sampled galaxies, the respective drawbacks of both codes, and the restriction of  coverage to the optical regime. The measured properties are compared to those reported in the COSMOS2020 catalogue, showing a good agreement. We have released the catalogue of measured properties in tandem with this work.
    }


   \titlerunning{PAU Survey: Galaxy stellar population properties with narrowband data}
   \authorrunning{B. Csizi et al.}

   \maketitle


\section{Introduction}
The stellar content of galaxy populations offers important insights into galaxy evolution and formation, as the physical properties can provide a direct indicator of evolutionary paths for tracing star formation histories and metal enrichment. At the same time, the physical properties of galaxies are predictive with respect to effects relevant to cosmological studies of galaxies, particularly for their clustering bias and the degree of the intrinsic alignment of a galaxy's shape relative to the local tidal field \citep{Benson2000, Li2006, Joachimi2013, Samuroff2021}. It is therefore vital to determine the distribution and the evolution of stellar populations to constrain models of galaxy evolution and answer current cosmological questions. 

The common technique used to obtain galaxy stellar population properties from photometric data is spectral energy distribution (SED) fitting \citep[for a review, see e.g.,][]{Conroy2013}. This process requires  three main ingredients, namely: the observational data, a physical model of the galaxy population, and a method of statistical inference to connect them. The fluxes or apparent magnitudes of observed or simulated galaxies are related not only to the underlying stellar population, but also depend on redshift, the emission from gas and dust in the interstellar medium (ISM), the history of star formation, and the dust attenuation of light along the line of sight. A template-based fitting process hence has to account for these contributions, which has the potential to introduce a large number of degrees of freedom into the routine, increasing consequently the complexity of the estimation.

The optimization of the SED fitting parameter space and a statistically powerful and efficient approach is vital for the characterization of galaxy populations in large-scale surveys. Therefore, a collection of panchromatic SED codes have been developed based on stellar population synthesis (SPS) to forward-model the galaxy population and fit a spectrum motivated by the physical processes that account for the observed spectral distributions \citep{Walcher2011}. These codes employ a variety of different frameworks or templates, as well as various Bayesian methods, such as Markov chain Monte Carlo sampling or $\chi^2$ minimization on a model grid \citep{Conroy2013, Thorne2021, Pacifici2023}. 

These SED fitting codes work by fitting real galaxy fluxes with ones generated from template SEDs. The higher the resolution of the photometric data, the more reliable will be the galaxy properties estimated via SED fitting. Additionally, the characteristics of the models generated by the fitting code influence the quality of the estimation, depending on their ability to describe the plethora of observed galaxies accurately. Different initial mass functions (IMFs) or dust attenuation laws for instance, as well as assumptions about the stellar metallicity, may introduce a difference in stellar mass estimates at a  $\sim 0.2$\,dex level \citep[see e.g.,][]{Speagle2014, Bernardi2018} or bimodality in the stellar mass distribution \citep{Mitchell2013}. Previous tests on mock catalogs by \cite{Mejia2017} determined an improved accuracy in physical parameter estimation from the Javalambre-PAU Astrophysical Survey (J-PAS) narrowband data in comparison to broadband fluxes for passive galaxies only. Additionally, studies with PAUS mocks showed high capability to recover galaxy color-redshift relations \citep{Manzoni2023}.

Narrowband surveys can be a powerful niche between spectroscopy and photometry for this task. While wavelength-resolved spectroscopic data only comes at a detriment to target selection biases, flux-limitation, and long integration times, broadband photometric surveys lack spectral resolution. However, an increased number of filter bands with narrow bandpasses can deliver a quasi-spectroscopic energy resolution with high accuracy photometric redshifts, while the imaging approach ensures time efficiency and thus a large sample size, which are of paramount importance for statistical cosmological research and the creation of a global model of the galaxy population. One such narrowband survey is the Physics of the Accelerating Universe Survey (PAUS) conducted at the Obervatorio del Roque de los Muchachos on La Palma, Spain \citep{Benitez2009, Marti2014, Serrano2023}. The survey uses 40 uniformly spaced narrowband filters with a full width at half maximum (FWHM) of 12.5\,nm in the optical and achieves photo-$z$ precisions up to $\sigma(z)/(1+z)\sim 0.0057$ for $i_{\text{AB}}<22.5$ over a volume of 0.3\,(GPc/h)$^3$ \citep{Eriksen2019, Alarcon2021, Soo2021, Cabayol2023, Navarro2023}.  

Previous medium- and narrowband surveys such as ALHAMBRA \citep{Diaz-Garcia2015}, miniJPAS \citep{GonzalezDelgado2021}, and J-PLUS \citep{Sanroman2019} have demonstrated the constraining power of selected SED codes in the analysis of simulated and observed data. Studies on PAUS photometries have shown the estimation capabilities for galaxy properties \citep{Tortorelli2021} and the improved detection of spectral features such as the 4000\,\AA{} break \citep{Stothert2018, Renard2022}. Nevertheless, the gain of narrowband photometry for the estimation of physical properties in comparison to $ugriz$ data alone has not been conclusively quantified yet and many widely used SED fitting codes still remain to be tested. The degeneracy of galaxy properties such as metallicity, age, and dust parameters lead to correlated influences on the SED that cannot be broken by the spectral forward-model \citep{Worthey1994, Sawicki1998, Bell2001, Papovich2001}. A quasi-spectroscopic energy resolution in the optical (average $\mathrm{R}\sim 65$), could be able to improve the constraints on galactic SFHs due to the capture of spectral features that trace galaxy properties, such as the 4000\,\AA{} break or the H$\beta$ index as stellar evolutionary trackers to infer ages \citep{Lee2000}, in addition to, for example, abundance ratios from emission lines to estimate the metal enrichment of the stellar population \citep{Tremonti2004}. 

In this work, we employ narrowband data from PAUS and broadband data from the Canada-France-Hawaii Telescope Legacy Survey \citep[CFHTLS,][]{Cuillandre2012, Gwyn2012} to determine galaxy stellar population properties in the PAUS-COSMOS field with two SED fitting codes, namely \cigale \citep{Burgarella2005, Noll2009, Boquien2019} and \prospector \citep{Leja2017, Johnson2021}. We apply the SED fitting routine to a sample of PAUS-observed galaxies, fitting both the narrowband and broadband combined as well as the broadband data alone, to determine the distribution of properties, quantify the benefit of the PAUS narrowband filter set, and identify drawbacks and advantages of the two codes. 

The two SED fitting codes have a range of hyper-parameters that can be tuned by the user. Therefore, in order to identify the best set of hyper-parameters that can be later applied to data, we generated a set of synthetic PAUS-like galaxies simulated with \prospector and the underlying  Flexible Stellar Population Synthesis (\fsps) framework \citep{Conroy2009, Conroy2010, ben_johnson2022}. The knowledge of the true values of simulated galaxy properties allows us to adjust the hyper-parameter space and benchmark the two codes against one another. The tuned parameter space is then used to measure stellar masses, star formation rates (SFRs) and mass-weighted ages for observed galaxies in the COSMOS field. The results are then compared to the COSMOS2020 catalog \citep{Weaver2022}, to which our sample is cross-matched by RA and DEC.  

Section \ref{sc:Obs_data} describes the PAUS and CFHTLS observational data used in our work. Section \ref{sc:Methods} presents in detail the two SED fitting codes \cigale and \prospector. Furthermore, we also provide a description of how we generate the  library of mock galaxies and the priors. In Sect. \ref{sc:alpha_results}, we tune the hyper-parameter of \cigale and \prospector by fitting mock galaxy photometry and compare the results between the two codes. Section \ref{sc:obs_results} uses the tuned hyper-parameter space to measure the stellar population properties of observed galaxies in the COSMOS field with narrowband+broadband, as well as broadband only, and compares them against COSMOS2020 catalogue properties.
Throughout this work, we assume a flat $\Lambda$CDM cosmology with $\Omega_m = 0.32$, $\Omega_\Lambda = 0.68$, and $H_0=67.4\,\mathrm{km\,s}^{-1}\,\mathrm{Mpc}^{-1}$ \citep{Planck2020}.

\section{Observational data}
\label{sc:Obs_data}

\begin{figure}
\includegraphics[width=\columnwidth]{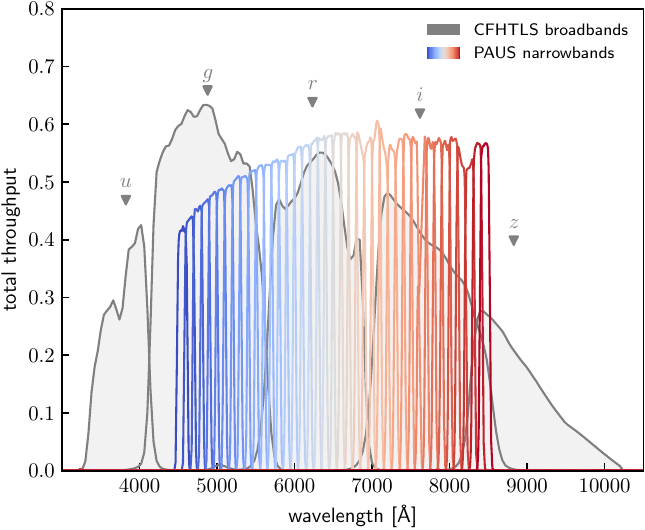}
\caption{Total throughput (including the effects of atmosphere, telescope optics, filter design and detector assembly) of the 40 PAUS narrowband filters (colored) spanning the wavelength range from 4550\,\AA{} to 8450\,\AA{} and the CFHTLS $ugriz$ broadband filters (gray).}
\label{fig:PAUS_CFHTLS_filters}
\end{figure}

This section describes the observational data analyzed in this work. The PAUS filter set contains 40 equally spaced slightly overlapping narrowband filters with a FWHM of 12.5\,\AA{} in the spectral range from 4550\,\AA{} to 8450\,\AA. The total throughputs of the filters, including atmospheric effects, as well as telescope, filter, and detector effects, such as the CCD quantum efficiency, are shown in Fig. \ref{fig:PAUS_CFHTLS_filters}. PAUCam has 18 red-sensitive full depletion CCDs, of which only the 8 central ones are used for narrowband imaging due to vignetting and image curvature. Its filters were commissioned to the prime focus of the \textit{William Herschel} Telescope and they cover a total FOV of $\sim$1\,deg$^2$ at an average pixel scale of 0.265" \citep{Benitez2009, Castander2012}. We use fluxes, flux errors and and photometric redshifts from the PAUS+COSMOS v0.4 flux/photo-$z$ catalogs hosted on CosmoHub 
\citep{Alarcon2021}.

The additional broadband photometry was taken from the final data release of CFHTLS \citep{Cuillandre2012}. This survey has been conducted at Mauna Kea, Hawaii, using MegaCam \citep{Boulade2003} and covers 175\,deg$^2$ in $ugriz$ bands with sub-arcsecond seeing at a pixel scale of 0.187". Figure \ref{fig:PAUS_CFHTLS_filters} shows a plot of the total throughput of PAUS narrowbands and the complementary CFHTLS broadband filters.

Both these surveys have common coverage in the COSMOS field ($\hat{=}$ CFHTLS D2 field), which is one of the most widely observed fields in astrophysics. Originally investigated by the Cosmic Evolution Survey \citep[COSMOS,][]{Scoville2007}  today there are multi-wavelength data available over a $\sim$2\,deg$^2$ surface centred at RA = 150.12 deg (10:00:28.6) DEC = +2.21 deg (+02:12:21.0) (J2000) near the celestial equator. More recent COSMOS catalogues of improved photometric redshifts and physical property estimates using a broader wavelength baseline including narrow- and medium band filters have since  been released \citep{Laigle2016, Weaver2022}, the latter of which is used in this work for comparison to our results. We performed the analysis on the common sample of both surveys, consisting of 25,491 galaxies up to redshift $z\sim 2$ and $i_\mathrm{AB}\sim 23.5$, all of which have observations in all 40 PAUS bands and the five CFHTLS $ugriz$ bands. 

\section{Methods}
\label{sc:Methods}
This section describes the SED fitting codes we employed for estimating galaxy stellar population properties from photometric data, namely \cigale and \prospector, focusing on their different evaluation approaches for inferring SEDs and the corresponding population properties using stellar population synthesis (SPS). We also describe the method we used to create simulated galaxy populations using \prospector with non-parametric star formation history with independent star formation rates over several time bins. This library contains synthetic PAUS and CFHTLS magnitudes and associated errors, as well as ground truth physical properties, which can then be utilized for parameter space optimization and the general testing of the SED fitting accuracy with narrowband photometries.

\subsection{\cigale}
The Code Investigating GALaxy Emission (\cigale), introduced in its newest \texttt{python} implementation by \cite{Boquien2019}, is a modular SED code covering emission from X-ray and FUV wavelengths up to the FIR and radio regime. Once a set of models for the desired contributions to the galactic SED has been chosen and a range of values for the free parameters has been set, \cigale computes the grid of these models and evaluates them on the data based on their $\chi^2$ value, which is calculated according to
\begin{equation}
    \chi^2 = \sum_j \frac{  
        \left[F_{\lambda_j}^{ \text{obs}} - \sum_{k,l,m} \alpha_{k,l,m}f_{\lambda_j}(\Psi_k, \zeta_l, \xi_m)
        \right]^2}{\sigma_j^2}.
\end{equation}
Here, $F_{\lambda_j}^\text{obs}$ is the observed flux in the $j^{\text{th}}$ band, $\sigma_j^2$ is the corresponding uncertainty of the observation, and $\alpha_{k,l,m}f_{\lambda_k}$ are the model fluxes created from a star formation history $\Psi$, a metal enrichment $\zeta$ and additional contributions from for instance dust or nebulae $\xi$. \cigale draws  simple stellar population (SSP) spectra from the Bruzual \& Charlot BC03 stellar evolution synthesis templates \citep{Bruzual2003} given an initial mass function (IMF) and a set of discrete metallicity values. These spectra are then combined with a parametric star formation history (SFH), attenuated by a flexible attenuation curve as well as corrected for emission from ionized gas and interstellar dust. \cigale takes the maximum likelihood model from the customized grid with a standard likelihood $\mathcal{L} = \exp(-\chi^2 / 2)$ as the resulting SED with its associated properties. For the BC03 templates, \cigale provides low- and high-resolution spectra, with the latter facilitating emission line recovery but with an increased run time. Version 2022.1 of the code is used throughout this project. 

\subsection{\prospector}
While the approach employed by \cigale can be computationally fast for low-dimensional problems, the grid size grows exponentially with the number of fitted parameters, thus resulting in high computational requirements for a complex parameter space. \prospector, on the other hand, finds the best-fit SED with Markov chain Monte Carlo (MCMC) sampling from an initial prior space using the Metropolis-Hastings algorithm implemented within the \texttt{emcee} package \citep{Foreman-Mackey2013}. \prospector draws hyper-parameter samples, creates observed-frame SEDs using the \fsps framework, generates mock photometry from the model SED and evaluates the log-likelihood $\log\mathcal{L}(D|\theta, \alpha)$, where D is the data and $\theta, \alpha$ are the model parameters and nuisance parameters. For our purposes, we employ a modified log-likelihood that consists of two separate likelihood terms, one for the narrowband and one for the broadband data, as: 
\begin{equation}
    \log\mathcal{L}_c = 0.8 \,\log\mathcal{L}_\text{NB} + 0.2\,\log\mathcal{L}_\text{BB}.
\end{equation}
Given the larger number of narrowbands and their higher spectral resolution, we assigned a greater weight to the narrowband (NB) likelihood with respect to the broadband one.

Limitations of the code mainly arise from the size or shape of the parameter priors, as for example large uniform priors provide a space that is too uninformative or takes too long to reliably cover with a reasonable amount of computing time, while narrow Gaussians can be too tight for a large sample of various galaxy types. \prospector provides as output the MCMC chains for all free model parameters of the fitting model, which constitute probability density functions (PDFs) that can be then used to infer more statistically reasonable properties. This can be achieved by exploration of the posterior space to for instance break degeneracies or analyze multiple peaks in the posterior, which lead to multiple possible physical property values for the respective galaxy SED fit. If only the best-fit model is desired, the maximum a posteriori (MAP) value can be determined from the chains. The current build v1.1 is used in this work.

\subsection{\palpha simulated stellar populations}
\label{sc:palpha}

To calibrate the hyper-parameters of \cigale and \prospector and use them on real data, we generated mock photometry using the \prospector code itself. We generate a library of 10\,000 synthetic galaxies with $0.1<z<1.0$ and $ i_\mathrm{AB} < 22.5$. These ranges are dictated by the PAUS limiting magnitude and the coverage of the bulk of the PAUS redshift distribution. We used the \palpha-model described in \cite{Leja2017}, \cite{Leja2019} to create the photometric data and the associated set of physical properties. This setup uses a non-parametric SFH, a two-component \cite{Charlot2000} curve for modeling the dust attenuation separated based on contributions from stellar birth clouds and the diffuse ISM optical depths (with the latter having a modified wavelength scaling that accounts for the UV bump, see \citealt{Noll2009}, \citealt{Kriek2013}), and a dust emission model from \cite{Draine2007} and \cite{Draine2014}, along with nebular emission using the $\textsc{Cloudy}$ templates \citep{Ferland2017}. We first drew 50\,000 samples from the prior space described in Table \ref{tab:simulations_priors} and then we randomly selected 10\,000 sources having $i_\mathrm{AB} < 22.5$.

In the \palpha model \citep{Leja2019}, galaxy stellar metallicities are drawn from the \cite{Gallazzi2005} mass-metallicity relation (MZR). For each draw from the stellar mass prior, a stellar metallicity value is sampled from a clipped normal distribution that is characterized by the means and standard deviations of the corresponding mass bin. $\sigma$ is thereby defined as the difference of the 84$^\mathrm{th}$ and 16$^\mathrm{th}$ percentiles. For consistency, we assume a \cite{Chabrier2003} shape for the IMF throughout this work. 

\begin{table}
\caption{\palpha parameters and priors}
\centering
\begin{tabular}{ll}
    \hline 
    Parameter & Prior distribution  \\
    \hline
    \hline
    Redshift $z$ & $c\mathcal{N}(\mu=0.6, \sigma=0.25, \min=0.1, \max=1.0)$ \\
    $\log(M_*/M_\odot)$ & $\mathcal{U}(\min = 9, \max = 12.5)$  \\
    $\log(U)$ & fixed at $-2$ \\
    $\log(Z_\mathrm{gas}/Z_\odot)$ & $\mathcal{U}(\min = -2, \max = 0.5)$ \\
    $f_\mathrm{bol}^\mathrm{AGN}$ & $\log\mathcal{U}(\min = 10^{-5}, \max = 3)$ \\
    $\tau_\mathrm{AGN}$ & $\log\mathcal{U}(\min = 5, \max = 150)$ \\
    $n$ & $\mathcal{U}(\min = -1, \max = 0.4)$ \\
    $U_{\min}$ & $\mathcal{U}(\min = 0.1, \max = 15)$\\
    $\gamma$ & $\mathcal{U}(\min = 0.0, \max = 0.15)$\\
    $q_\mathrm{PAH}$ & $\mathcal{U}(\min = 0.1, \max = 7.0)$\\
    $\hat{\tau}_{\lambda, 2}$ &  $c\mathcal{N}(\mu=0.3, \sigma=1, \min=0, \max=4)$ \\
    $\hat{\tau}_{\lambda, 1} / \hat{\tau}_{\lambda, 2}$ & $c\mathcal{N}(\mu=1.0, \sigma=0.3, \min=-3.3, \max=3.3)$ \\
    \hline
\end{tabular}
\tablefoot{Parameters and their priors from which we sample our input for the photometric simulations using the \palpha-model:\ $\log(U)$ is the ionization parameter, $f_\mathrm{bol}^\mathrm{AGN}$ is the AGN fraction of the bolometric luminosity, $\tau_\mathrm{AGN}$ is the optical depth of the AGN dust torus, $n$ is the slope modifier of the Calzetti attenuation curve \citep{Calzetti2001}, $U_{\min}$ is the minimum dust emission radiation field, $\gamma$ is its illuminated fraction, $q_\mathrm{PAH}$ is the fraction of PAHs, $\hat{\tau}_{\lambda, 2}$ is the diffuse dust optical depth, and $\hat{\tau}_{\lambda, 1}$ is the birth cloud optical depth in the two-component model from \cite{Charlot2000} . In this table,  $\mathcal{U}$ specifies a uniform prior that is bound by $\min$ and $\max$ values, $\mathcal{N}$ designates a Gaussian prior characterized by a mean $\mu$ and a standard deviation, $\sigma;$ finally, $c\mathcal{N}$ is a clipped Gaussian prior that can be used to truncate a specific parameter to exclude, for example, sampling from negative values.}
\label{tab:simulations_priors}
\end{table}

The SFH priors for the non-parametric SFH are set over seven age bins distributed according to \palpha up to the age of the universe at the respective redshift. Therein, the ratios of SFRs $x_{n, n+1}$ in adjacent bins are modeled by independent priors with a Student's $t$-distribution:
\begin{equation}
    s_t(x_{n, n+1}, \nu) = \frac{\Gamma\left(\frac{\nu+1}{2}\right)}{\sqrt{\nu\pi}\,\Gamma
    {\left(\frac{\nu}{2}\right)}} 
    \left(
    1 + \frac{(x_{n, n+1}/\sigma)^2}{\nu}\right)^{-\frac{\nu+1}{2}}
,\end{equation}
with $\nu=2, \sigma=0.3$. Using these ratios and the total stellar mass formed, \prospector computes the SFRs for each age bin, from which a recent SFR over a specified time range (usually 100 Myr) can be calculated to be able to later compare with the SFR estimates from \cigale. 

\begin{figure}
    \centering
    \includegraphics[width=\columnwidth]{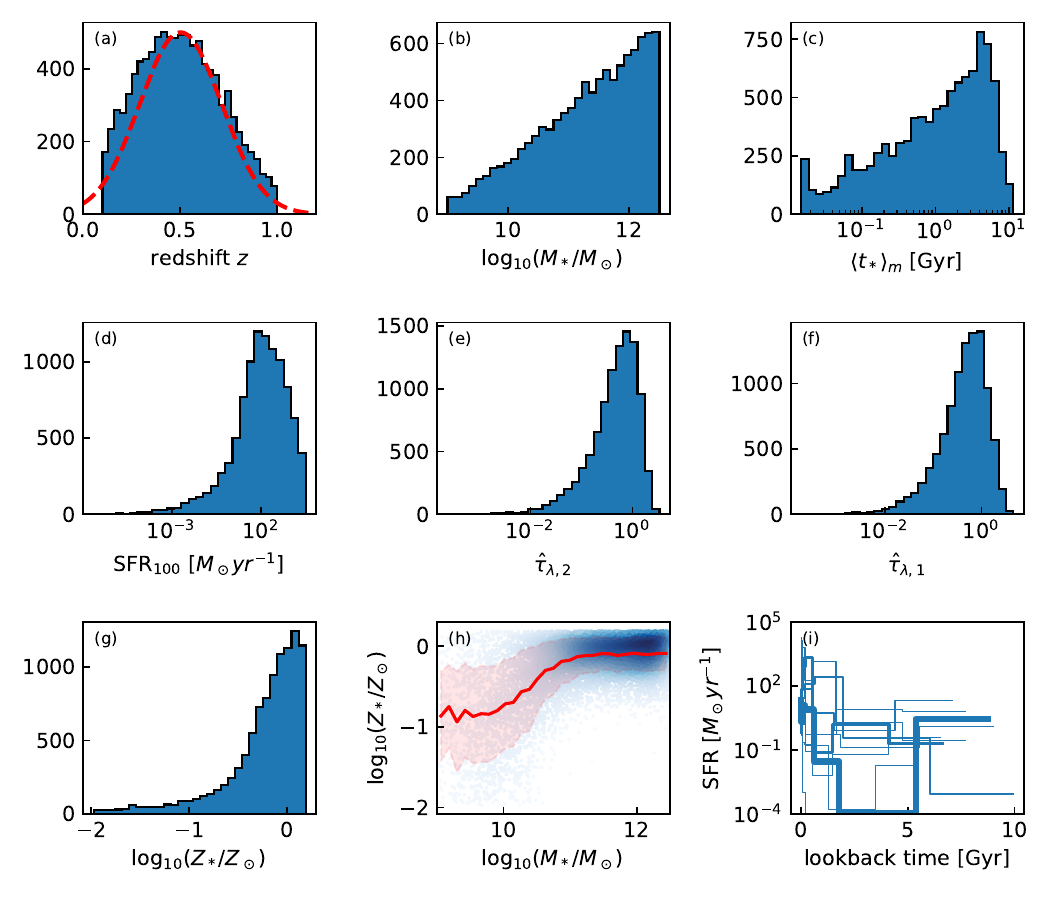}
    \caption{Prior distributions of selected parameters used for the generation of mock spectra and photometries for (a) redshift $z$ using the PAUS redshift distribution (red); (b) logarithm of the stellar mass; (c)  mass-weighted stellar age $\left\langle t_*\right\rangle_m$; (d)  SFR over the last 100\,Myrs; (e)  stellar metallicity $Z_*$; (f-g) the ISM and BC optical depths, $\hat{\tau}_{\lambda, 2}$, $\hat{\tau}_{\lambda, 1}$; (h)  mass-metallicity relation adopted from \cite{Gallazzi2005}, and (i) random sample of ten of the non-parametric SFHs.}
    \label{fig:alpha_library_hist}
\end{figure}

Figure \ref{fig:alpha_library_hist} shows the distributions of the parameters we are most interested in for the scope of this work. We also show the mass-metallicity relation assumed for the joint prior sampling of stellar mass and stellar metallicity, as well as a few non-parametric SFHs from the sample. Note that due to the i-band magnitude cut, the distributions do not necessarily still follow a uniform or Gaussian PDF, as visible for example in the $\log(M_*/M_\odot)$ histogram.

For each of the 10\,000 prior draws, we generated the galaxy SED with \prospector and we obtain synthetic PAUS and CFHTLS model magnitudes. The SED fitting routines compute the best-fitting parameters taking into account the photometric errors. Therefore, to realistically mimic the SED fitting measurement of real data, we assigned errors to the galaxy model magnitudes. The CFHTLS photometric errors are assigned by modeling the magnitude errors as function of magnitude relation using a Gaussian process regressor. To do this, we first split the CFTHLS COSMOS field $ugriz$ data into thin magnitude bins and compute mean and standard deviations of the magnitude errors in each bin. Then, the resulting data was used to train a Gaussian process regressor. The latter was then used to predict the magnitude errors based on the input model magnitudes. The same approach can be implemented for PAUS filters, by using the Flagship-PAU catalogue from Castander et al. (in prep.), \cite{Cabayol2023}, which includes photometry of \textasciitilde 130\,000 simulated galaxies in the PAUS bands with $i_\mathrm{AB}<23$. 

\section{Optimization with simulated galaxies}
\label{sc:alpha_results}

The following section presents the results of SED fitting on the simulated photometric library. We use simulated galaxies to optimise the hyper-parameter space for \cigale and \prospector. Afterwards, we present the results of both codes on the simulations for a set of galaxy physical properties that we deem relevant for this study. 

\subsection{Parameter Space Selection}

\renewcommand{\arraystretch}{1.1}
\begin{table}
\caption{\cigale parameter space}
\centering
\begin{tabular}{p{0.45\columnwidth}|p{0.45\columnwidth}}
    \hline 
    Parameter & Grid values  \\
    \hline
    \hline
    \multicolumn{2}{l}{\textit{BC03 SSP templates}} \\
    \hline
    IMF & 1 (Chabrier) \\
    Metallicity $Z_*$ & 0.0001, 0.0004, 0.004, 0.008, 0.02, 0.05 \\
    \hline
    \multicolumn{2}{l}{\textit{delayed-$\tau$ SFH}} \\
    \hline
    e-folding time $\tau_*$ [Gyr]  & 0.1, 0.5, 1, 2, 4, 6, 8, 10, 12 \\
    Age $\langle t_*\rangle $ [Gyr] & 0.1, 0.5, 1, 2, 3, 4, 5, 6, 7, 8, 9, 10\\
    Burst/quench age $t_{\text{bq}}$ [Gyr] & 0.01, 0.1, 0.5 \\
    SFR ratio of burst/quench $r_\text{SFR}$ & 0.0, 0.1, 0.2, 0.6, 1.0, 3.0, 10.0 \\
    \hline
    \multicolumn{2}{l}{\textit{Nebular emission}} \\
    \hline
    Ionization parameter $\log(U)$ & -2.0 \\
    Gas-phase metallicity $Z_\text{gas}$ & 0.001, 0.014, 0.041 \\
    \hline
    \multicolumn{2}{l}{\textit{Charlot \& Fall dust attenuation}} \\
    \hline
    ISM $V$-band attenuation $A_V^\text{ISM}$& 0.001, 0.01, 0.2, 0.4, 0.6, 0.8, 1.0, 1.5, 2.0, 2.5\\
    $A_V^\text{ISM} / (A_V^\text{BC}+A_V^\text{ISM})$ & 0.3, 0.5, 0.7, 1.0 \\
    Power-law slopes $n_\text{ISM}$, $n_\text{BC}$ & -1.0, -0.5, 0.4\\
    \hline
    \multicolumn{2}{l}{\textit{Draine \& Li dust emission}} \\
    \hline
    PAH fraction $q_\text{PAH}$ & 3.90 \\
    Minimum radiation field $U_{\min}$ & 1.0 \\
    Illuminated fraction $\gamma$ & 0.01 \\
    \hline
\end{tabular}
\tablefoot{Overview of the module and parameter selection for \cigale. Unspecified parameters of the respective model are fixed at their default values. A list of parameters for each module is available in the appendix of \protect\cite{Boquien2019}.}
\label{tab:parameters_cigale}
\end{table}

The comparison between the input properties of the \palpha simulations and the SED fitting procedure applied on simulations allows us to fine-tune the selection of models and associated parameters and priors for \cigale and \prospector.

In doing so, we decided to keep the redshift fixed to its true input value. A similar choice was performed also on real data, where the redshifts are fixed to the values quoted in the PAUS photo-$z$ catalogue described in detail in \cite{Alarcon2021}. Keeping the redshift fixed allows us to reduce the dimensionality of the problem and remove degeneracies between the redshift and the galaxy physical properties.

In Table \ref{tab:parameters_cigale}, we list the grid of hyper-parameter values for the \cigale SED fitting routine. These values have been obtained by finding the right parameter combination that returns the highest degree of consistency between the recovered properties and the input ones. The hyper-parameters are utilised throughout this work, including the estimation of real galaxy properties. In \cigale, we used the BC03 simple stellar population templates and a delayed-$\tau$ shape with optional late star-burst or quench for the SFH \citep{Lee2010, Madau2014}.  For a schematic overview of this SFH's parameters, we refer to, for example, Fig. 1 in \citealt{Manzoni2021}. Furthermore, we employed a \cite{Charlot2000} dust attenuation curve, a parameterized dust emission templates from \cite{Draine2007} and the {\scshape Cloudy} nebular emission templates. The main differences to the simulated galaxies lie in the stellar templates, MILES \citep{Sanchez-Blazquez2006} vs BC03 for Prospector and \cigale, respectively, and in the use of the delayed-tau SFH instead of the non-parametric one. \cigale does not contain the latter option in the same manner, as there is a non-parametric SFH, but without the possibility of constraining SFR ratios between adjacent bins. 

For the SED fitting with \prospector, we use the same \palpha setup that had generated the synthetic galaxies. The only difference is in the stellar metallicity prior which is chosen to be uniform, $\mathcal{U}(\min = -2, \max = 0.5)$. We chose this distribution due to our limited knowledge about the metallicities of the observed galaxies as we intend to apply the same model to simulations and observations. Thus, this broad uninformed prior should not be able to introduce an additional bias into the estimation. We use the MILES empirical stellar spectral library and the MIST isochrones \citep{Choi2016}.

\subsection{Results with \cigale}

\begin{figure}
    \centering
    \includegraphics[width=\columnwidth]{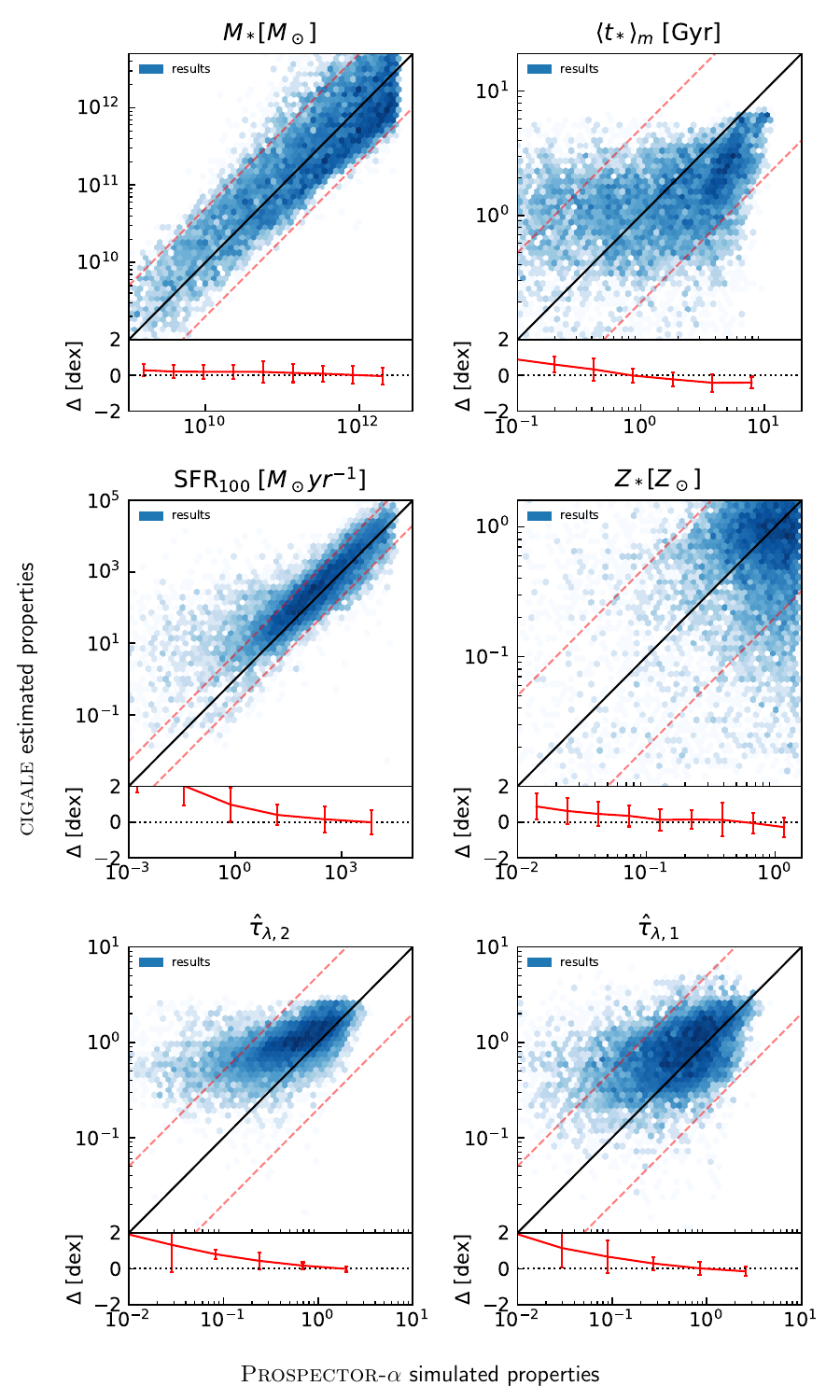}
    \caption{Hexbin plots showing the ground truth from the \palpha stellar population simulations on the x-axis versus the MAP value from the \cigale $\chi^2$ analysis, for stellar mass $M_*$, mean mass-weighted stellar ages \agemw, SFR over the last 100\,Myrs, stellar metallicity $Z_*$ and the optical depths of the two-component dust attenuation model $\hat{\tau}_{\lambda, 2}$ and $\hat{\tau}_{\lambda,1}$. The red points and errorbars show the median and standard deviation for several bins. Below each panel, we show mean the residuals and their standard deviations over ten bins covering the parameter range.}
    \label{fig:cigale_on_fsps_sims}
\end{figure}

The hyperparameter grid from Table \ref{tab:parameters_cigale} has been determined as the one providing the highest degree of consistency with the input simulated data. We show in Fig. \ref{fig:cigale_on_fsps_sims} the Bayesian maximum a posteriori estimates from \cigale for six SED properties in relation to the true input values of the \palpha mock photometries. The properties shown are the stellar mass, $M_*$, the mean mass-weighted stellar ages \agemw, the SFR over the last 100\,Myrs, the stellar metallicity, $Z_*$, and the optical depths of the two-component dust attenuation model $\hat{\tau}_{\lambda,1}$ and $\hat{\tau}_{\lambda,2}$ for the contributions from birth clouds and the ISM, respectively. Additionally, below each 1:1 relation, we show the residuals $r_i = \langle p^C_i \rangle - \langle p^\alpha_i\rangle$ and their standard deviations $\sigma^p_i$ for 10 bins covering the parameter range, where $\langle p^C_i \rangle$ and $\langle p^\alpha_i\rangle$ are the mean values in the $i^{\text{th}}$ bin from \cigale and the simulated library, respectively. The respective bin means are also shown directly in the plot.    

The results show that the input stellar masses are well recovered by \cigale for most of the sample, as shown by the residuals being consistent with zero over the whole mass range. The absolute values of the residuals tend to increase towards small stellar masses. The results are not surprising since stellar masses have been shown to be quite stable against different choices of the SFH and spectral resolution, leading to robust estimates from even simple SPS models  \citep{Pforr2012}. 

The input SFRs averaged over the last 100\,Myrs, instead show  a different behavior. They are well recovered for input SFRs down to roughly $10^0 M_{\odot}/yr$, but they depart from the 1:1 relation for lower values of the input SFR. In particular, for a fixed input SFR, \cigale tends to overpredict the SFR with respect to the simulations, with residuals up to 2-3 dex for input SFRs smaller than $10^{-2}\,M_\odot$\,yr$^{-1}$. The results imply that our \cigale setup is not able to reliably recover the range below SFR$<10^0\,M_\odot$\,yr$^{-1}$.

\begin{figure}
    \centering
    \includegraphics[width=\columnwidth]{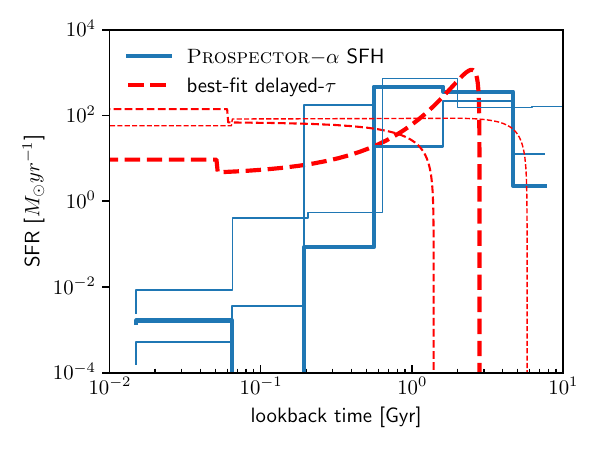}
    \caption{Example comparison of non-parametric \palpha SFHs (blue) and the corresponding \cigale parametric delayed-$\tau$ model best-fit (red) for three galaxies where the current SFR estimation error exceeds two orders of magnitude. Corresponding objects have the same line thickness in the plot.}
    \label{fig:example_bad_sfh}
\end{figure}

We attribute these large residuals to the different SFH models, as the SFR recovery from spectra using SPS tends to largely depend on the selected SFH model as well the dust priors due to the dust-age-metallicity degeneracy \citep{Sawicki1998, Papovich2001}. To compare the star formation histories between the two data sets, we show in Fig. \ref{fig:example_bad_sfh} example SFHs for strong outliers where the difference between input SFR and fitted value exceeds two orders or magnitude, so $|\Delta(\text{SFR})|>10^2$. 

Figure \ref{fig:example_bad_sfh} depicts three such comparisons. Clearly, the SFHs differ heavily by their current and integrated star formation rate, as well as its temporal development and the maximum age of the galaxy stellar population. These large model differences naturally translate to the best-fit SPS parameters. The SPS framework incorporated in \cigale, using BC03 templates as simple stellar populations of different metallicities, differs from the generation of SEDs in our simulations with \palpha, leading to potential inconsistencies due to the usage of different isochrones and stellar spectral libraries. Additionally, the delayed-$\tau$ model cannot achieve the same level of flexibility as the non-parametric SFH from \palpha, as seen in Fig. \ref{fig:example_bad_sfh}.

As a result, mass-weighted ages show a less consistent behavior with respect to stellar masses and SFRs. The turn-over seems to be at 1 Gyr: galaxies with input ages smaller than 1 Gyr have systematically over-predicted ages, while input ages larger than 1 Gyr lead \cigale to under-predict the mass-weighted age, although with smaller residuals with respect to small input ages. The  mean mass-weighted stellar age is heavily dependent on the individual SFR ratios of the \palpha SFH continuity bins, while the functional form of the delayed-$\tau$ cannot account for multiple changes on the overall rate of star formation, for instance an initial starburst followed by a later quench once the galaxy passes through the green valley. Nevertheless, we would have expected a better agreement of the recovered ages with respect to the input ones due to the addition of the PAUS data, as for example the 4000\,\AA{} break, a strong age indicator, is within our baseline for the majority of the sample galaxies. While the PAUS filters do not cover the break in the rest-frame, it shifts into the available wavelength range at approximately $z=0.13$, meaning that $>95\,\%$ of our simulated sample have narrowband coverage at the 4000\,\AA{} break. 

Mean stellar metallicities of the galaxy population have residuals from the one-to-one relation that are consistent with zero for the range half-solar to super-solar metallicities. Low input metallicity galaxies are instead scattered in the whole parameter space so they are not recovered at all. One reason for difficulties in metallicity estimation is the sparsely sampled metallicity grid within the BC03 templates accessible in \cigale, which can lead to an uninformative posterior and thus inaccurate property estimates. Still, even at $Z\approx Z_\odot$, the scatter is significant. This shows that there are other influences on the fitting accuracy such as the age-dust-metallicity degeneracy, making it difficult to obtain reliable estimates on either of these properties without strongly constrained priors. Moreover, at low-$Z$, SSPs are complicated in general, with also MILES-based models only being able to model old ages reliably \citep{Vazdekis2010, Coelho2020}. 

As ages and metallicity in particular are strongly correlated with emission lines and the 4000\,\AA{} break, we tested if the usage of the high-resolution BC03 templates improves the recovery. We however find only a slight improvement on the metallicity with no effect on age estimation. Therefore, we choose to employ the low-resolution stellar spectra throughout this work in order to reduce computational requirements. 

The optical depths, $\hat{\tau}_{\lambda,1}$ and $\hat{\tau}_{\lambda,2}$, of the two-component dust model have residuals consistent with zero in the range 0.2-2, while they heavily differ at smaller input values. This is not surprising since age, dust and metallicity are degenerate and one parameter influence the other.

\subsection{Results with \prospector}

\begin{figure}
    \centering
    \includegraphics[width=\columnwidth]{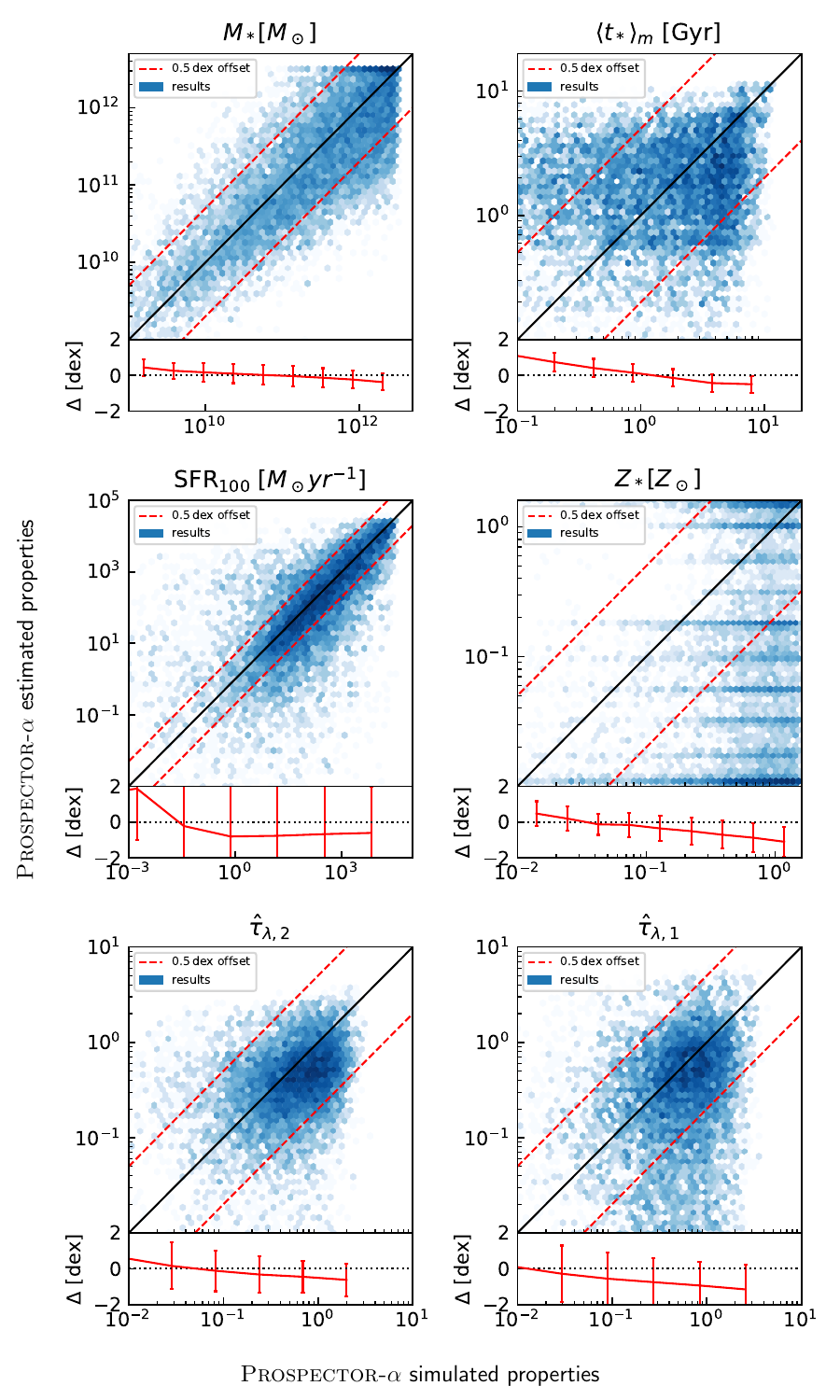}
    \caption{Hexbin plots showing the ground truth from the \palpha stellar population simulations on the x-axis versus the MAP value from t \prospector, for stellar mass, $M_*$, current SFR, mean mass-weighted stellar ages, \agemw, stellar metallicity, $Z_*$ and the optical depths of the two-component dust attenuation model, $\hat{\tau}_{\lambda, 2}$ and $\hat{\tau}_{\lambda,1}$.}
    \label{fig:prospector_on_fsps_sims}
\end{figure}

Applying \palpha to the set of mock photometries created with the model itself should allow to consistently recover reasonable posteriors and allow the fine-tuning of the MCMC parameters, such as the number of walkers and iterations. 
We run \palpha on the simulated photometries to recover the model hyper-parameters and estimate the fitting accuracy. In Fig. \ref{fig:prospector_on_fsps_sims} we again show the same comparison of the maximum a posteriori SED estimates to true simulation values for the \prospector code. 

We find robust estimates for stellar masses and SFRs, which in this case is expected due to the same star formation histories used for the generation of the data and the forward-modeling of the fitted SED. Dust parameters are also recovered accurately on a 0.5\,dex level for the majority of the sample galaxies, albeit with some objects where the birth cloud dust attenuation is not reliably recovered. The overall intrinsic scatter from the 1:1 relation apparent for every free parameter is a consequence of fits where the MCMC chain is not fully converged yet after the set number of iterations, as well as parameter degeneracies and the simulated uncertainties of the model photometry. 

Our setup, however, struggles to estimate accurate stellar metallicity posteriors, mainly for older populations with high $Z_*$, where the estimated values seem to be mostly clipped to the edges of the uniform prior space. We attribute this to the prior differences during the simulation and the sampling phase, where we once chose a joint mass-metallicity prior and later a uniform distribution for sampling. Increasing the number of likelihood calls with more MCMC iterations or more walkers can help to alleviate this problem, as we restricted these hyper-parameters due to computational limitations. We show in Appendix \ref{apdx:posteriors} an example fit with an increased iteration count, which can lead to substantial shift of the best-fit parameters. Additionally, and most likely due to the age-metallicity degeneracy, $\langle t_* \rangle_m$ is not estimated well for a substantial subset of mock objects, where especially the ages of young stellar populations are overestimated by up to one order of magnitude. This then also influences the metallicity estimates and hinders an accurate measurement of both of these parameters without a joint prior. 

The main caveat of the \prospector code is the run time of each likelihood calls. With a large prior space of 16 free parameters, many of which being broad uniform priors, a large number of walkers and iterations is needed to have a converging posterior with high fidelity accurate SED fitting and parameter estimation. Using 2048 iterations of 48 walkers, which is our selection throughout this work, leads to fitting times of up to 24 hours per galaxy on a single core. For future applications to large galaxy datasets, improved sampling algorithms have to be implemented that reduces the computational requirements. 

Additionally, we note that the reliance on only the MAP value of the posterior distribution is usually not always sufficient, mostly due to degeneracies of the parameters. Looking at some full posteriors from our estimates, one can identify multiple peaks with a high likelihood that where found exploring different regions of a joint dust-age-metallicity prior. We show an example of such a posterior distribution in Appendix \ref{apdx:posteriors}.

\section{Stellar population properties}
\label{sc:obs_results}
Using our selection of fitting hyper-parameters and \cigale grid values, we show the results from both codes on the sample of 25\,491 galaxies in the COSMOS field. We complete two runs for \cigale on the catalog, once using CFHTLS $ugriz$ data only and once with additional 40 PAUS narrowband fluxes. For \prospector, we only run the code once with narrowband+broadband data, due to the required $\sim$ 500k CPU hours for one run on the full sample. We first show the overall distributions of our parameter estimation as well as the robustness of the inferred properties. We also perform a comparison between the two codes and the COSMOS2020 catalog to determine the predictive power of the PAUS narrowband filter set in relation to previous studies with broad filter baselines that also include narrowband photometries.

\begin{figure}
    \centering
    \includegraphics[width=1.0\columnwidth]{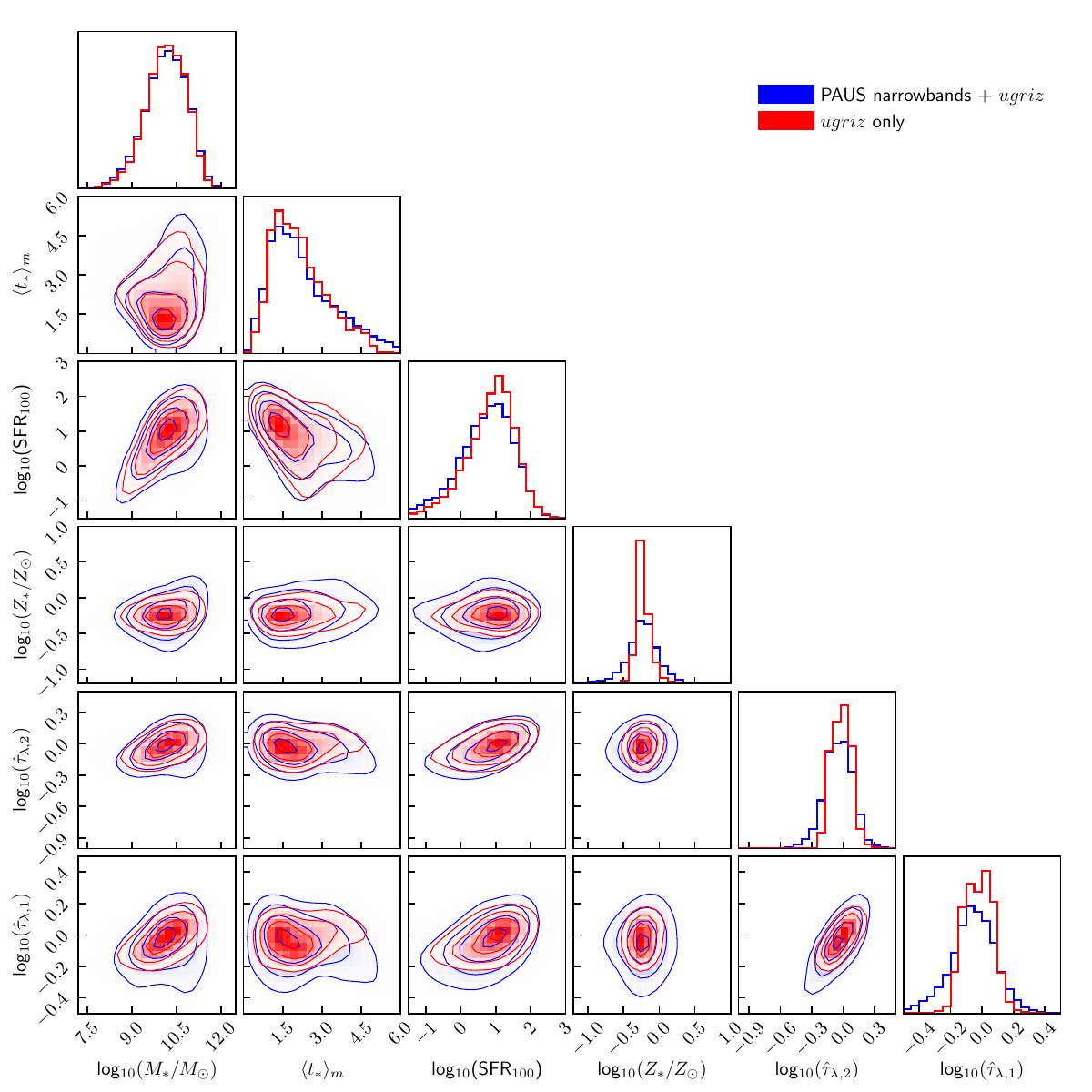}
    \caption{Distribution of MAP values of the selected properties as determined by \cigale: stellar mass, mass-weighted stellar age, SFR over 100\,Myrs, stellar metallicity and optical depths of ISM and BC dust components. Blue shows the results from using narrowband and broadband data, red depicts the properties from using CFHTLS $ugriz$ alone.}
    \label{fig:cigale_on_obs}
\end{figure}

\subsection{SED fitting with CIGALE}

In Fig. \ref{fig:cigale_on_obs}, we show the distributions of the inferred physical properties obtained with \cigale on our catalogue of 25\,491 galaxies in the COSMOS field. The blue contours and histograms show the distributions resulting from running the SED fitting routine on the combined PAUS (narrowband) and CFHTLS (broadband) dataset, while the red contours were determined by using the broadband $ugriz$ data alone. 

We first note that the stellar mass, which usually represents the most robust property inferred by SED fitting, is similarly constrained by either using only the broadband or the combination of narrowband and broadband data.
The other properties, especially metallicity and dust component optical depths,  show that with the addition of the narrowband, we can capture a broader range of physical values with respect to the sole use of the broadband. Since these physical properties are strongly correlated to distinct spectral features, the net effect of an increased photometric resolution manifests in the form of a more diverse distribution. Broadband filters alone tend to estimate metallicities that only slightly differ from $Z_*=Z_\odot$, as well as ages towards the high-end tail of the mass-weighted age distribution and low optical depths $\hat{\tau}_{\lambda, 2;1}$ of the dust attenuation model. This can be interpreted as a result of the large impact of the prior choice and size on these parameters in the broadband-only case, resulting in a narrower posterior.

Nevertheless, the effect of the narrowband data is not as strong as expected, where most of the sample galaxies are still similarly fitted with reasonable property estimates from the underlying SPS forward-model using $ugriz$ data alone. However, we see that narrowbands do not lead to to a different region of the parameter space with respect to the use of broadband data only, which indicates that the model is performing in the same manner with either setup. Additionally, we note that the overall coverage of the prior ranges set by the \cigale parameter space from Table \ref{tab:parameters_cigale} is largely diminished in the posterior distribution. This means that either the priors are too wide and the true observations do not cover the ranges set from the \palpha model, or the sparse grid leads to an incomplete sampling of the high-dimensional parameter space. This is facilitated by the degeneracies of certain model properties. We fill further explore this after analyzing the \prospector results and the comparison with COSMOS2020. 

To validate the inferred properties robustness we plot the $M_*$-SFR relation for five redshift bins up to $z<1.0$. This relation reveals the main populations of galaxies, namely the star forming main sequence (SFMS) and the red sequence of older, non-star-forming quiescent galaxies. There exist multiple empirical laws for the SFMS that have made use of various SFR tracers to obtain a robust formulation of the redshift dependence of this sequence \citep{Popesso2023}. Previously, more variable laws over larger redshift ranges have been proposed, for example by \cite{Thorne2021}, where the SFR as a function of stellar mass is given by 
\begin{equation}
    \log_{10}\;\text{SFR} = S_0 - \log_{10} \left[
    \left(\frac{M_*}{M_0}\right)^\alpha + \left(\frac{M_*}{M_0}\right)^\beta \right], 
\end{equation}
with $S_0$, $M_0$, $\alpha$, and $\beta$ as empirical parameters that depend on redshift. \cite{Popesso2023} recently presented a new parametrization of the star forming main sequence as a function of time, $t$, by combining previous studies to a common calibration following:
\begin{equation}
    \log_{10}\;\text{SFR} = a_0 + a_1t -
    \log_{10} (1+(M_*\, 10^{-a_2-a_3t})^{-a_4}), 
\end{equation}
where $a_0,a_1,a_2,a_3$ are the model parameters, for which we use the best-fit values from the original paper. 

\begin{figure}
    \centering
    \includegraphics[width=\columnwidth]{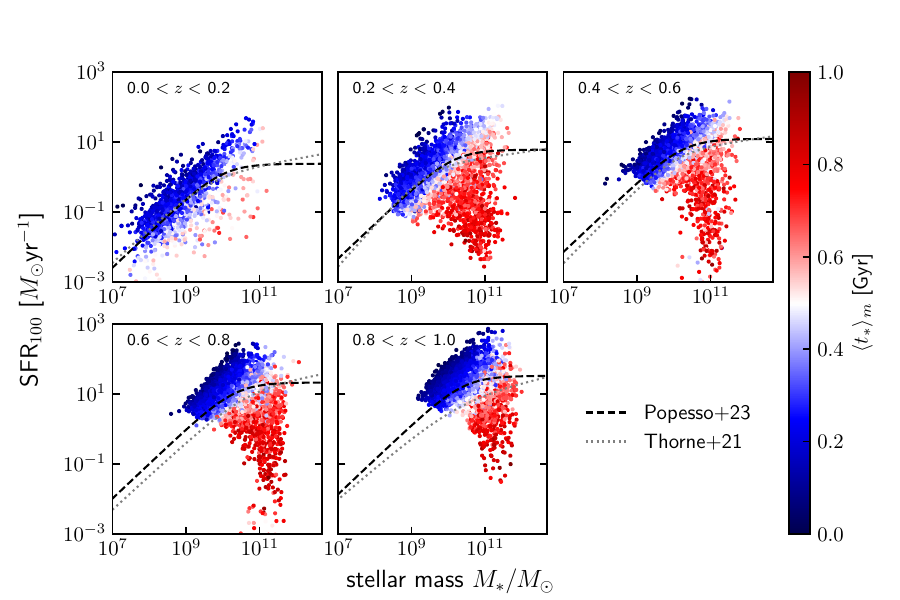}
    \caption{SFR as function of $M_*$ for five redshift bins in the range $0<z<1.0$, colored by the mean mass-weighted stellar age. The dashed line depicts the respective Popesso SFMS law and the dotted gray line the Thorne SFMS law.}
    \label{fig:cigale_fsms}
\end{figure}

Figure \ref{fig:cigale_fsms} shows the $M_*$-SFR relation for the different redshift bins, color coded by their mean mass-weighted stellar ages. It can be seen that the main population of young, star forming galaxies (SFMS) matches with the empirical laws mostly for the low-$z$ subset, while a constant offset to higher SFRs with respect to the empirical formulation can be observed. This behavior is similar to the overestimation bias measured with \cigale on the simulations and likely has the same source, that is the simplified \cigale SFH model that can heavily reduce the constraining power on related properties. Overall however, older, red and quiescent galaxies exhibit lower SFRs and lie below the SFMS, with intermediate galaxies where star formation has started quenching populating the green valley between the main clouds \citep{Gallazzi2017, Kalinova2021}. 

\begin{figure}
    \centering
    \includegraphics[width=\columnwidth]{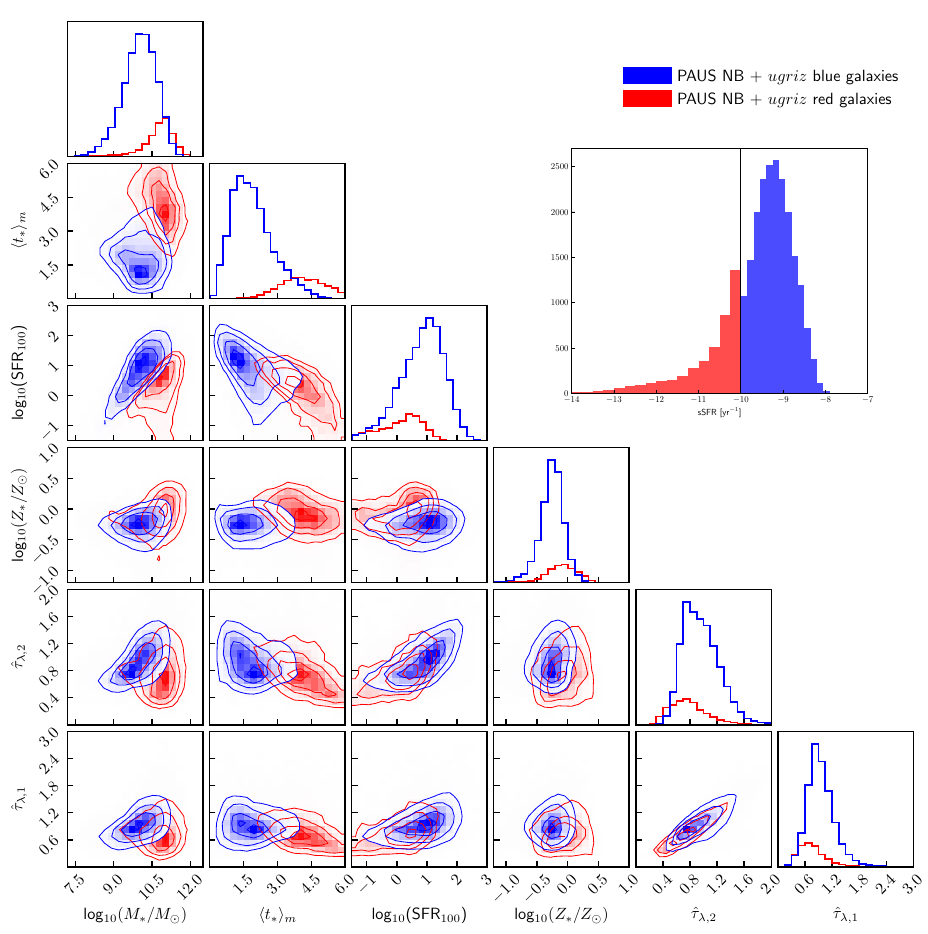}
    \caption{Histogram of sSFR values estimated from \cigale and corner plot of estimated properties. Blue contours and histograms indicate the main sequence of star forming galaxies, red depicts the red sequence of old galaxies The solid line in the histogram shows the cut at $=10^{-10}$\,yr$^{-1}$, separating the blue and red galaxy populations.}
    \label{fig:cigale_res_sep}
\end{figure}

\begin{figure*}
    \centering
    \includegraphics[width=1.0\columnwidth]{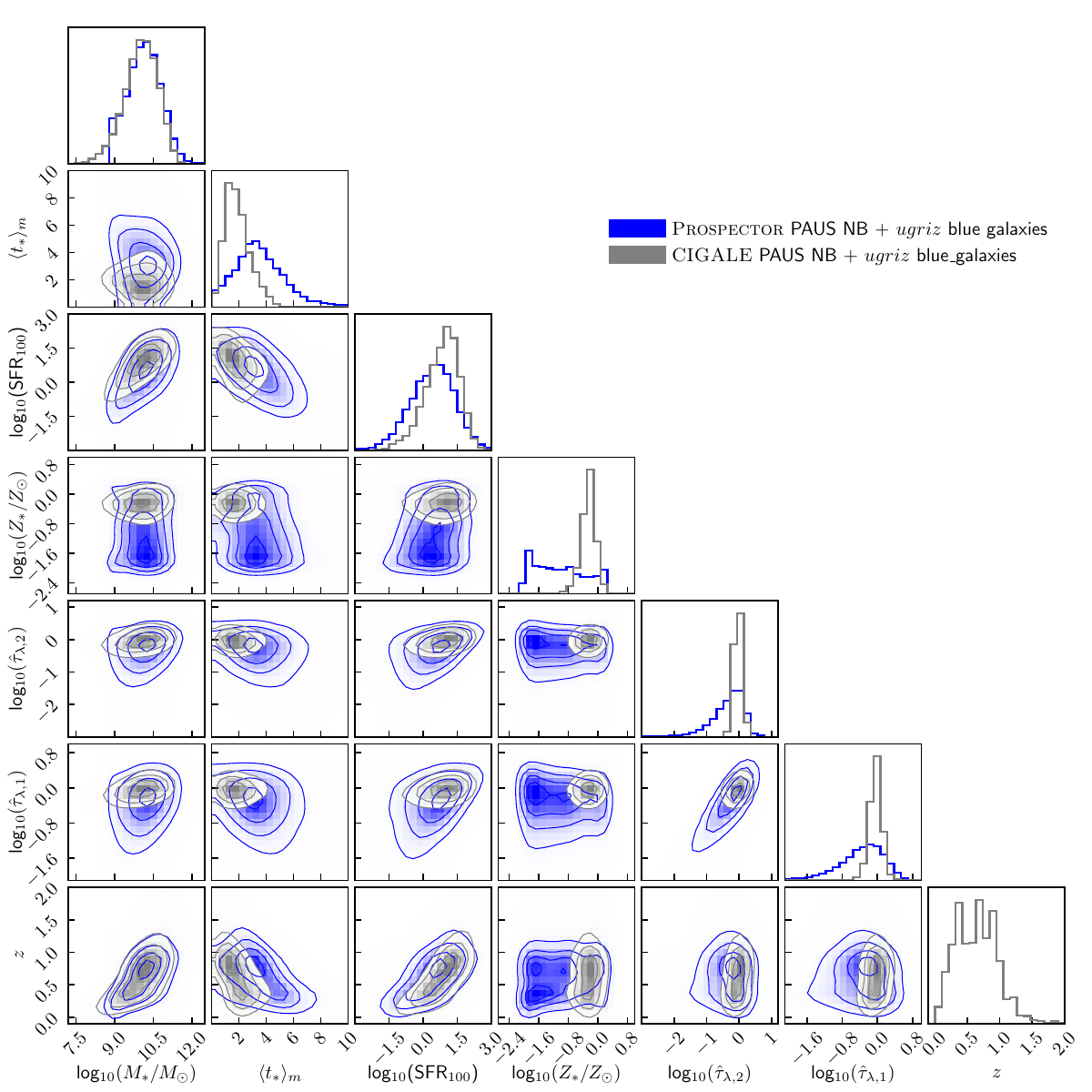}
    \includegraphics[width=1.0\columnwidth]{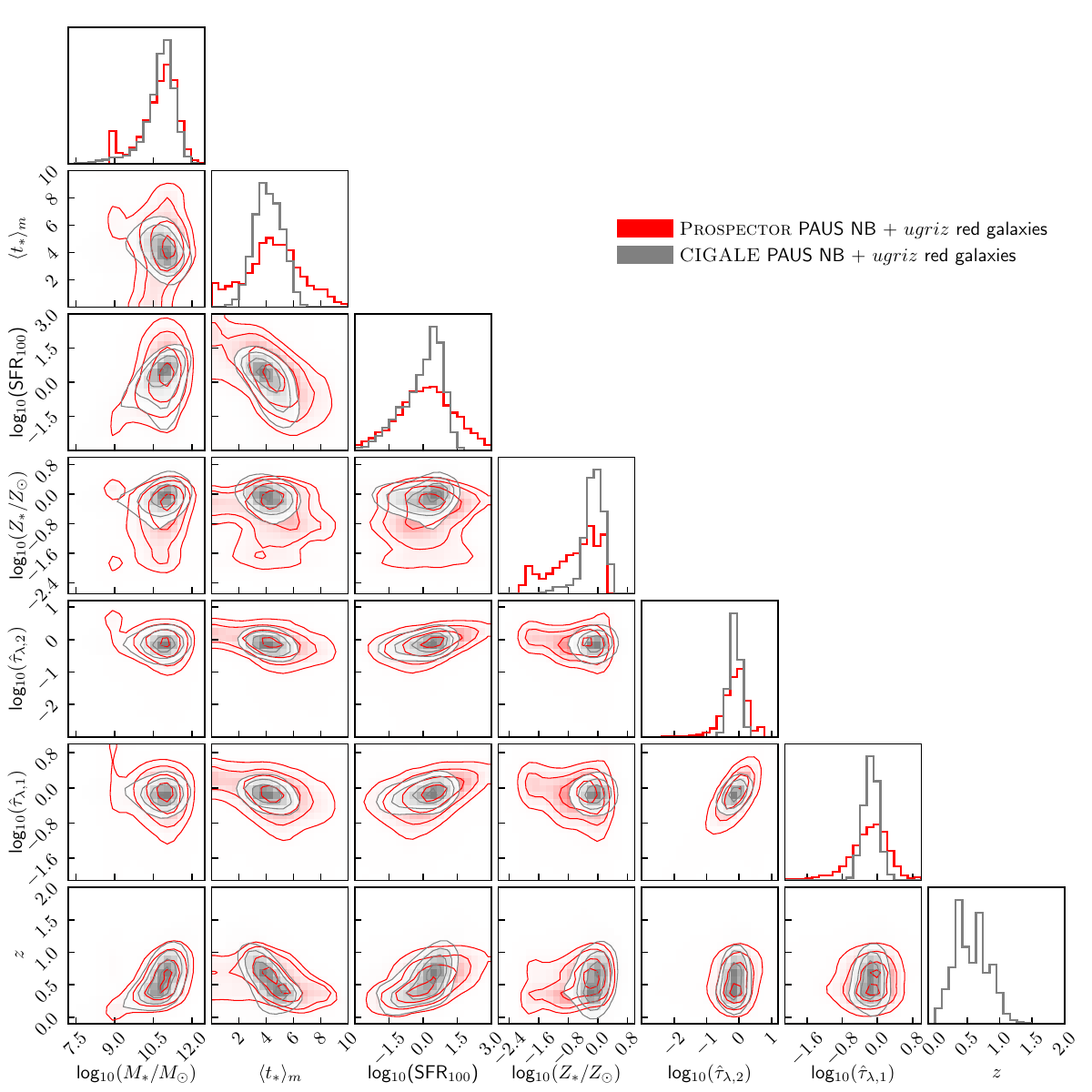}

    \caption{Distribution of MAP values of the selected properties as determined by \prospector: stellar mass, mass-weighted stellar age, SFR over 100\,Myrs, stellar metallicity the optical depths of the ISM and BC dust components, all using both narrowband and broadband $ugriz$ data. We additionally show the redshift $z$. The two subplots depict the blue and red galaxy population after an sSFR cut, gray shows the respective populations from \cigale.}
    \label{fig:prospector_on_obs}
\end{figure*}

We go on to we separate the main galaxy populations by a cut of the specific star formation rate, $\mathrm{sSFR}=\mathrm{SFR}/M_*$. The threshold is hereby set as:
\begin{equation}
    \mathrm{sSFR}_\mathrm{quiescent} \leq 10^{-10}\,\mathrm{yr}^{-1} 
    < \mathrm{sSFR}_\mathrm{star-forming},
\end{equation} 
as previously used by \cite{Weinmann2006}, \cite{Salim2009}, \cite{LaraLopez2010}, \cite{Annunziatella2014}, \cite{Tortorelli2018}. Applying this cut to the \cigale output allows us to plot the distributions of properties for these two main populations. Figure \ref{fig:cigale_res_sep} shows again the distributions of estimated properties, now separated by their affiliation to either the red or blue galaxy population, respectively, as well as a histogram of the sSFR with the aforementioned cut. 

The mass-weighted ages of the red sequence peak at around 4\,Gyr, while the blue cloud has a mean age of around 1.5\,Gyr; their stellar masses show distributions peaked at $\sim 10^{11}\,M_\odot$ and $\sim 10^{10}\,M_\odot$, respectively.

It is apparent that we are able to recover the bi-modality of galaxy populations \citep{Baldry2006}. Galaxies identified as red, with $\mathrm{sSFR}\leq 10^{-10}\,\mathrm{yr}^{-1}$, are older, have higher mass, lower SFR, contrary to blue star forming galaxies that have higher SFR and $\tau$, are younger and thus have also a lower $M_*$. The bimodality is not as evident for the dust components and $\log(Z/Z_\odot)$, but we can still observed the younger population to have lower mean stellar metallicities and more attenuation by interstellar dust, as expected. 

\subsection{SED fitting with Prospector}

In this section, we analyze the PAUS narrowbands and CFHTLS $ugriz$ observational data by running the Prospector MCMC estimation. We also plot the property distributions of the maximum a posteriori SED of the chains, which is displayed in Fig. \ref{fig:prospector_on_obs}. In this figure, the galaxies are already separated by the aforementioned sSFR cut into blue and red populations and shown with the respective \cigale 2D distributions. We also plot the photometric redshift $z$, although this is not a free parameter of the fitting routine and just depicts the value from the PAUS catalog. 

Overall, we can clearly see that the distributions appear largely different from the \cigale results for some of the properties. For once, the mean age estimate for blue galaxies is much higher with \prospector, while it also finds best-fit values over a larger range of the age prior. Mass-weighted stellar ages there have values up to 10\,Gyrs from \palpha. This effect can be traced throughout the entire parameter space, where the \prospector overall property ranges strongly exceed the distribution widths from the previous \cigale estimates, with especially the dust components and metallicity exhibiting a more widely spread posterior space. Still, the means of the 1D histograms are similiar between both codes, for instance, for $\hat{\tau}_{\lambda, 2}$, $\hat{\tau}_{\lambda, 1}$.

The galaxies however do not show a pronounced mass-bimodality, with both populations covering the full range of values as opposed to the \cigale results. This could, however, be caused by the tight \palpha mass prior, which does not allow SED models with $M_* < 10^9\,M_\odot$, resulting in a pile-up at the prior edge. Still, we observe a clear separation in mass-weighted ages and SFRs, however again not as clearly for the metallicity and $\hat{\tau}_{\lambda, 2}$, $\hat{\tau}_{\lambda, 1}$. 

The size of the posterior is here mainly related to the size of the prior, as \cigale is not able to sample outside of its prior and thus cannot cover the same parameter ranges as \palpha due to the restrictive parameter space. We note however that increasing the grid values of \cigale for example for the dust components did not yield better results. Hence, the differences are likely rooted in the sampling techniques and the different SPS models -- mainly the star formation history complexity which, aside from age and SFR, especially influences metallicity and dust attenuation. The box-like shape of the \prospector metallicity distribution suggests that the accessible value range is either still not wide enough, that the prior is not informative enough (or the number of MCMC iterations is too low), making it hard to sample high-quality constraints, or that the photometry is not able to constrain this property accurately. Given the results for $Z_*$ on the simulation, we favor the latter explanation, given the inherent dust-age-metallicity degeneracy. 

\begin{figure}
    \centering    
    \includegraphics[width=\columnwidth]{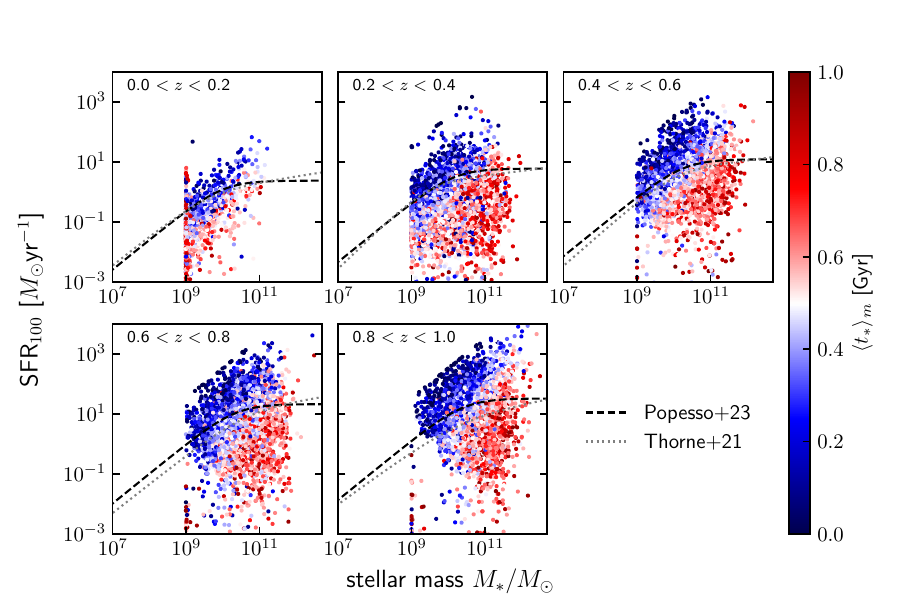}
    \caption{SFR as function of $M_*$ for five redshift bins in the range $0<z<1.0$, colored by the mean mass-weighted stellar age as found from \prospector. The dashed line depicts the respective Popesso SFMS law and the dotted gray line the Thorne SFMS law.}
    \label{fig:prospector_fsms}
\end{figure}

To validate our findings we again plot the SFR-$M_*$ relation over the specified redshift bins in Fig. \ref{fig:prospector_fsms} and compare it to the \cigale results. First of all, the clear cut-off in the mass prior is visible, as no galaxies with $M_*<10^9\,M_\odot$ appear. Overall, the subplots paint a similar picture to the \cigale analysis, where the parametric law fits well at lower redshifts, but increasingly breaks down at higher photo-$z$. Still, we observe a clear separation by mass-weighted stellar age and roughly follow the empirically expected SFR-$M_*$ relation. It is however clear that a broader mass-prior would be beneficial for the analysis of our dataset, thus favoring the \cigale mass estimates.

The discrepancies between \cigale and \prospector for $Z_*$,  ages, $\hat{\tau}_{\lambda, 2}$ and $\hat{\tau}_{\lambda, 1}$ are interesting to observe, as they indicate the strong dependence of SED fitting results on prior choices and SPS models, where just a more complex SFH model might strongly change the parameters of the best-fitting spectrum. The overall distributions of SFRs and stellar masses however are mostly similar. This is expected for stellar masses, however, this is not entirely natural for the SFR, given that it is directly related to the SFH. Since the \cigale parameter space was however chosen by analysis on \palpha simulations, a similar constraining power can be explained by the SPS model optimization process.

Given the similar measurement quality of the dust attenuation components for both codes on the simulations, the differences on the observations on these specific properties require further explanation. The distribution of \palpha priors, from which the simulated photometries where sampled, likely does not consistently cover the whole range of true parameters in the COSMOS field. \cigale and \prospector will not access SPS models with parameters outside of the fixed grid or the prior, respectively, that we matched towards the simulation priors. As both codes handle the likelihood maximization very differently, discrepancies may naturally arise for some galaxies, that can be mostly attributed to the age-dust-metallicity degeneracy, which are the parameters for which we observe the strongest differences. Besides, finding a low-$\chi^2$ fit is more complicated with many tightly spaced narrowband fluxes, which may lead to inefficient exploration of the prior space or strong degeneracies.

Broadband SED fitting quality studies might have avoided such issues due to simpler template fitting approaches or smaller filter baselines, where the likelihood of a model SED is more easily maximized given a low number of widely spaced filters. Previous narrowband results from for example J-PAS also attributed differences in spectra and SED fits of narrowband photometry to the signal-to-noise ratio (S/N) and the age-metallicity degeneracy due to a lack of corresponding indicators in the baseline \citep{Mejia2017}. The impact of this might not be as important for photo-$z$ estimates, but it would require further study on a broader range of SED codes to quantify the effect of SPS model choices. Still, we acknowledge that our \prospector results could demand a longer sampling process to cover the large high-dimensional prior space and thus might provide more similar results to \cigale with longer chains or more walkers. On the other hand, the latter does not allow a flexible SFH and a more sophisticated sampling method, which could facilitate a better exploration of the parameter space. Such techniques, such as MCMC, although infringe the code's applicability to next-generation surveys, where the amount of data requires faster techniques. In general however, one expects higher quality results from the MCMC fitting with a flexible non-parametric SFH prior, as in the \palpha model. Still, we do not know which of the results more closely matches the real values, as there exists no ground truth for observations and we thus cannot be certain if our priors tested on mocks are able to generate the real plethora of observed SEDs. A possible validation can be made through comparison with COSMOS2020, although we cannot be sure about the quality of the physical parameter estimates in the catalog either, due to the same limitations as with our setup. 

\begin{figure}
    \centering    
    \includegraphics[width=\columnwidth]{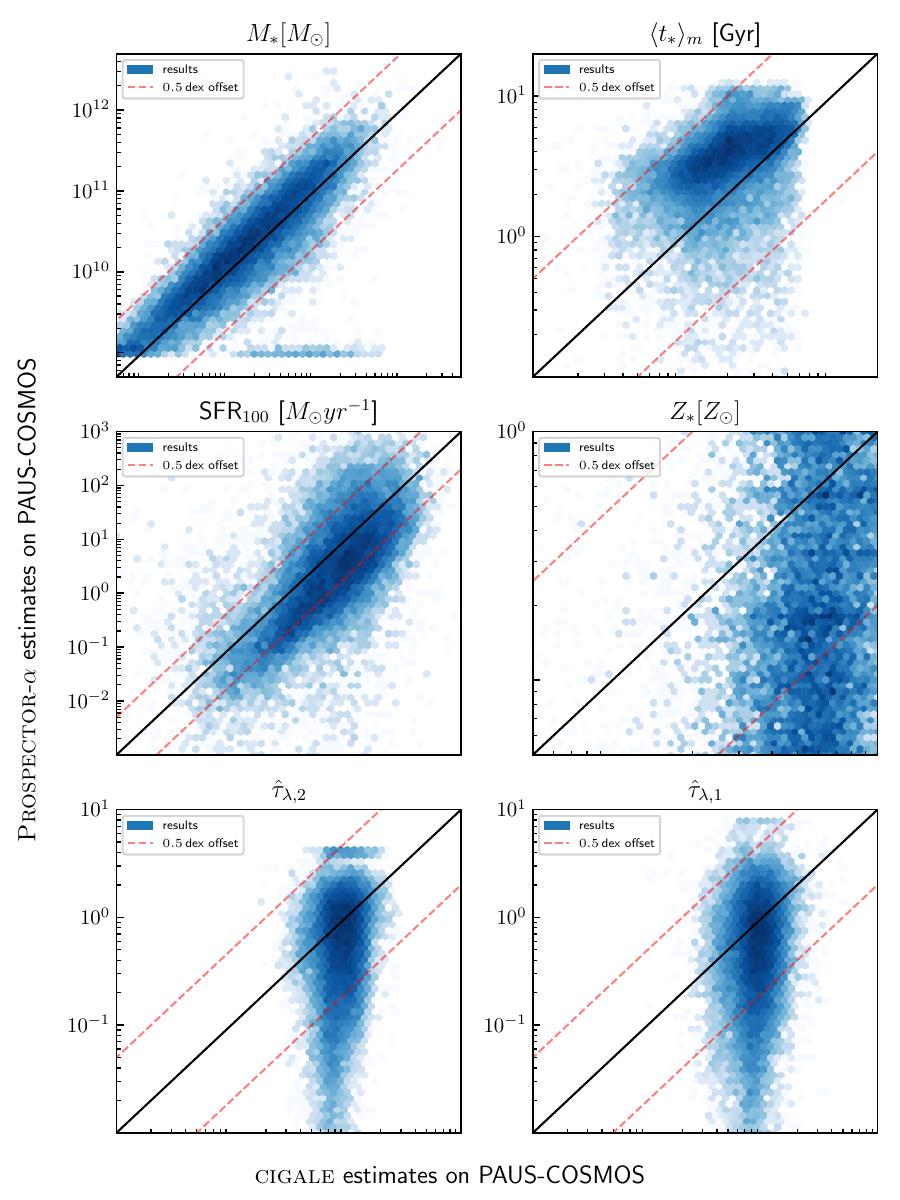}
    \caption{1:1 relations to compare the results on the PAUS-COSMOS observations with both codes. On the $x$-axis we plot the \cigale estimates on physical parameters, while the $y$-axis depicts the respective values from \prospector.}
    \label{fig:cig_prosp_1to1}
\end{figure}

Finally, we can also compare the results on a per galaxy basis to show the real discrepancies between both codes below the population level analysis (given in Fig. \ref{fig:cig_prosp_1to1}). As discussed earlier in this paper, the stellar mass estimates are robust within 0.5\,dex, aside from clipping at the prior edge with \palpha. Similarly, SFRs are constrained well, which is not usually the case from photometry, suggesting a powerful hyper-parameter setup. While ages also seem constrained better than expected given the two different SFH approaches, a 0.5\,dex level accuracy is large for estimates on a two orders of magnitude scale. Still, both codes find alike SFHs for a majority of the sample. The other properties on the other hand exhibit stronger discrepancies. Mainly the metallicity is not well constrained by the photometry as both codes produce extremely different values and distributions. For the dust parameters, \cigale does not find best-fit models over the whole grid size, while \palpha covers a larger range of values, leading to matching values only close to the mean of the distribution. Given the varying $Z_*$ and $\langle t_*\rangle_m$, it is a natural consequence that also the third contribution to the aforementioned degeneracy does not produce a congruent relation between the two setups. To remove this issue and break the degeneracy, at least partly, in the future, sampling via a joint prior that is calibrated on a spectroscopically observed mass-metallicity relation (as done for the creation of simulated SEDs with the \citealt{Gallazzi2005} prior), will be beneficial, although this is for now only possible for the \prospector code. 

Recent studies on the reliability of SED fitting techniques in general by \cite{Pacifici2023} tested the modeling uncertainties on different physical properties between different codes. They found that the stellar mass distribution is the most robust, with a higher uncertainty for SFH-related parameters (such as the SFR, especially at low SFR) and dust properties, due to their correlation. This relates well to our findings, where mostly the dust optical depth and metallicity distributions can vary strongly between the two setups. Therefore, modeling choices, as well as caution during interpretation of SFH, dust and metallicity parameters without spectroscopic information, are of paramount importance for SED fitting. 

Aside from the statistical framework, the prior selection, the likelihood function and the dust-age-metallicity degeneracy, the two setups also differ in their stellar template libraries (MILES for \prospector and BC03 for \cigale). Previous studies have tested the effect of varying spectral libraries with fixed fitting codes to test their respective contributions on physical property estimates. \cite{Coelho2009} and \cite{Dias2010} found that the choice of spectral templates (BC03, PEGASE-HR \citealt{Leborgne2004} and MILES) can have a significant impact on the age estimate in for example composite stellar populations of M32 or the Small Magellanic Cloud (SMC), without strongly affecting the overall metallicities derived from the integrated spectra. \cite{GonzalezDelago2019} on the other hand found a stronger effect of the stellar library choice (MILES, GRANADA \citealt{Martins2005}, STELIB \citealt{Leborgne2003}) on the metallicity, with typically 0.1\,dex and 0.3\,dex model-to-model dispersions for stellar ages and metallicities, respectively. This was however performed on stellar clusters, which can be characterized by a single age and do not require complex modeling a composite stellar populations over a wide range of ages. A recent work by \cite{Byrne2023} reported a stronger impact of the stellar spectral library on stellar population synthesis, where  line indices,  can particularly vary strongly between different templates, leading also to deviations in age and metallicity. Extensive studies over the impact on the assumptions of SED models are also outlined in Siudek et al. (in prep), who found that using different SSPs leads to a median absolute bias on a 0.25\,dex or 0.2\,dex level for the stellar mass and the SFR, respectively. This difference can be related to the incorporation of incorporation of thermally pulsating asymptotic giant branch (TP-AGB) stars in \cite{Maraston2010} models, which are not included in the BC03 templates. Overall, this shows that the observed differences between the two setups are not only caused by the different code properties, but also by the basis of the applied SPS process itself. In Appendix \ref{apdx:ssp-comp}, we plot a comparison of simulated composite stellar populations from \cigale and \prospector given different input SSPs and compare how such differences affect the simulated observed magnitudes in the PAUS and CFHTLS bands. We there show that even without influences from dust or gas models, the best-fit photometry from simulated spectra will vary at up to $\sim 0.2\,$mag level between our two setups for identical input physical parameters.

\subsection{Comparison with COSMOS2020}

To relate our results to previous findings with different codes and a larger photometric baseline, we compare the findings from \cigale and \prospector with our full dataset to the estimations from the COSMOS2020 catalogue calculated using the photo-$z$ code \textsc{eazy} \citep{Brammer2008}. We note here that the COSMOS2020 analysis incorporates >40 filters across a broad baseline, thus also including IR and UV data, but without a fine resolution within the optical such as PAUS. We again match the two libraries by RA and DEC and get 22.899 galaxies with common coverage. As the metallicity and dust extinction were fixed to solar/half-solar metallicity and $A_V=0.1$, respectively, and no delayed-$\tau$ or non-parametric SFH was used for the determination of COSMOS2020 properties, we can only compare stellar masses, stellar ages and the SFR. 

\begin{figure}
    \centering
    \includegraphics[width=\columnwidth]{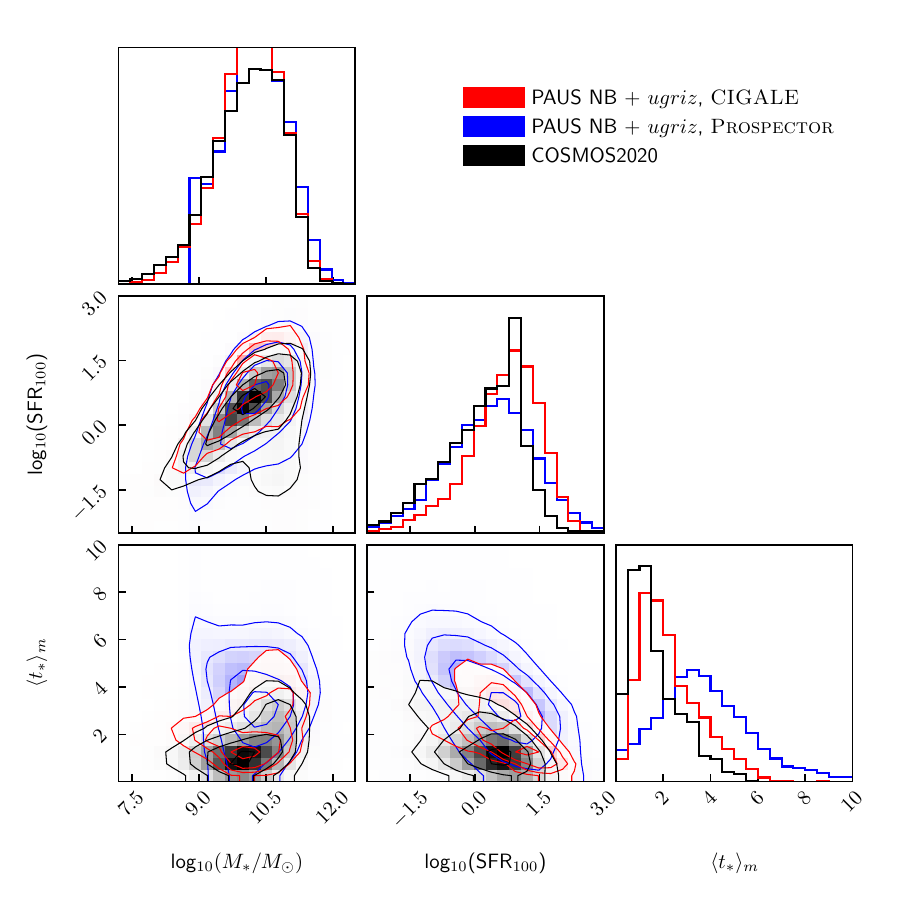}
    \caption{Comparison of stellar mass, age and SFR from our measurements with \cigale and \prospector and the \textsc{EAZY} estimates from the COSMOS2020 catalog.}
    \label{fig:comp-cosmos}
\end{figure}

\begin{figure}
    \centering
    \includegraphics[width=\columnwidth]{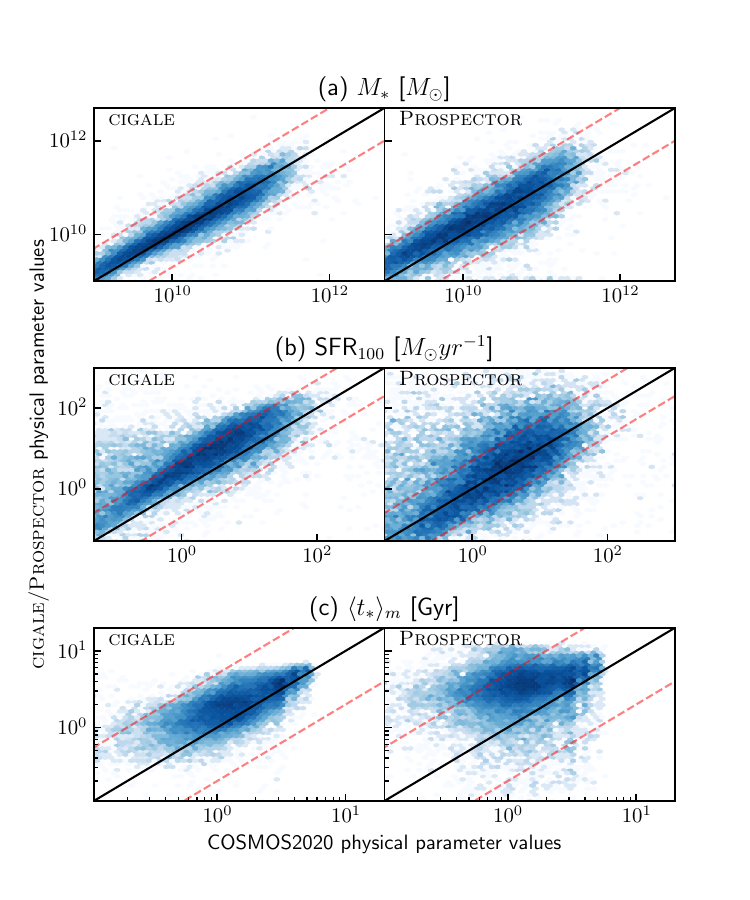}
    \caption{Comparison of physical parameter estimates between \cigale, \prospector and COSMOS2020. The $x$-axis shows COSMOS2020 results, while our estimates for each code are plotted on the $y$-axis.}
    \label{fig:comp-cosmos-1to1}
\end{figure}

Figure \ref{fig:comp-cosmos} depicts the overall distribution contours between the COSMOS2020 data, and our \cigale and \prospector estimates with narrowband PAUS and broadband CFHTLS observations on the matched sample. We find that the properties estimated with \cigale match more closely those from the COSMOS2020 catalogue, in particular for the stellar mass and the stellar ages, while Prospector better matches the SFR distribution of COSMOS2020. Even though COSMOS2020 finds a larger population of red sequence objects towards lower SFRs, the SFMS is mostly similarly recovered with both of our codes and within COSMOS2020. The ages, on the other hand, show similar  distributions between COSMOS2020 and \cigale; however, with a offset of around 1\,Gyr, where either COSMOS2020 underestimates the ages or \cigale overestimates them. As previously shown, the \prospector results on ages strongly differ from the other two samples, which we relate to the aforementioned different SFH treatment that allows overall higher ages with a more pronounced right tail of the distribution. 

Stellar masses, the most robust property from SED fitting, exhibits closely matching histograms, with the main difference being an overall more peaked histogram for \cigale and \prospector due to some galaxies having no mass estimate in the COSMOS2020 data set. Additionally, \prospector only has a uniform mass prior between $10^9\,M_\odot$ and $10^{12.5}\,M_\odot$, which explains the sharp cutoff and the distribution differences at the low-mass tail. Combined, we observed mostly similar SFR-$M_*$ planes for all samples, with COSMOS2020 and \prospector measuring more low-SFR red sequence galaxies compared to \cigale.

Again, we also show the 1:1 relations for both codes with respect to the COSMOS2020 property values in Fig. \ref{fig:comp-cosmos-1to1}. As expected, stellar masses are well recovered and have matching values in all three results. The SFR appears to be offset from the 1:1 relation with \cigale, with considerable scatter towards higher SFRs. The latter observation can also be made for \prospector, although the values here more closely match the COSMOS2020 findings for the majority of galaxies. Looking at the stellar ages, \cigale finds similar best-fit values for most of the objects, while the difference are increasingly large and biased towards higher ages with \prospector. This however is again mostly related to the SFH differences, as the COSMOS2020 fitting setup also relies on simple parametrizations of the star formation history, which manifest is largely different stellar ages when compared to non-parametric model fits. These results show the difficulties in determining mean stellar ages of galaxy stellar populations, even in the presence of high-resolution photometry of broad baselines, as it is the case for the COSMOS2020 data. 

\subsection{Differences to previous narrowband results}
Previous studies of SED fitting with narrowband data have been performed as part of the miniJPAS and J-PAS surveys, where the effect of an increase in photometric resolution and of a change of fitting code on the SED and physical property estimates was analyzed \citep{Mejia2017, GonzalezDelgado2021}. These studies found reduced biases in the physical parameter estimation with miniJPAS narrowbands with respect to broadband data. Moreover, they determined that the age-metallicity degeneracy is mostly hidden at broadband resolutions, only emerging at quasi-spectroscopic resolutions. While we cannot confirm a reduction in deviations between narrowbands and broadbands on these parameters, the larger parameter space indicates some degree of improvement due to the higher resolution.  

Moreover, we confirm that especially stellar masses are mostly independent of the applied SED model, with a robust estimate for all fitting codes. Similarly, we identify difficulties in the age recovery, as the miniJPAS results also found distribution mismatches between different codes for this property. On the other hand, we find a strong dependence of the metallicity on the SED code, which is not apparent in their results. We relate this though to the short sampling and the more flexible dust models applied, as well as the difference in stellar template libraries in our setups, which was not an issue for miniJPAS due to application of the latest version of the BC03 templates \citep{Plat2019} for all codes. 

Overall, such discrepancies between similar studies again emphasize how unstable physical property estimates even at similar resolutions can be. If the SPS model assumptions, stellar spectral libraries, or inferences frameworks differ, varying biases in the recovery of physical properties of stellar populations of galaxies have to be expected. This arises especially for ages, dust attenuation parameters and the mean stellar metallicity, given the difficulty in their estimation with clearly resolved emission lines or other spectral features. Recent work by \cite{Nersesian2024} also concluded that stellar populations properties from SED fitting with \textsc{Prospector} are dominated by the modeling approach even with deep photometric data with a broad baseline, including narrowbands. This, combined with the differences between our results and previous narrowband studies, indicates that the increased spectral resolution of current narrowband surveys does not yield an improved robustness of galaxy physical properties measurements, due to the large biases induces by the SPS modeling choices.

\section{Conclusion}
\label{sc:conclusions}
In this work, we tested SED fitting to quantify the effect of optical narrowband photometries from the PAU Survey on the predictive power of galaxy stellar population properties with modern SPS codes. We find that while the accessible range of parameter values is increased with the narrowband filters using \cigale, the overall difference in comparison with $ugriz$ data only is not especially pronounced. This can be caused by multiple reasons: non-flexible SPS models, inefficient parameter space coverage, likelihood maximization difficulties, or low narrowband S/N values. With \prospector, tests have to be performed to validate whether the metallicity issue might be caused by the rise of the age-metallicity degeneracy with increased spectral resolution, for example by comparing to broadband fitting only. 

Still, we find that the PAUS filter set, as well as the CFHTLS broadbands, both of which only cover the optical range and thus cannot rely on for example UV or IR bands for a more robust estimation of related properties, such as the SFR, are still able to recover previous measurements with larger baselines, that is COSMOS2020. With the majority of our sample covering a redshift range where the 4000\,\AA{} break lies within the regime of the PAUS bands, a more reliable SFH modeling is facilitated that allows robust SFR constraints. 

Furthermore, we show that the \palpha model is able to recover stellar masses and SFRs of the COSMOS sample, although with similar values compared to \cigale. This code provides moreover increased flexibility in its age range and can provide SEDs with more diverse dust and metallicity values. To test how much of this discrepancy compared to \cigale and COSMOS2020 depends on the PAUS filters will have to be further explored by \prospector fits on only $ugriz$ broadbands. As this however requires an extreme amount of computing resources for MCMC chain convergence per galaxy, the applicability to large data sets is heavily constrained. While we expect the \prospector code to be able to create more diverse SEDs with physically motivated SPS models, the broader range of age-dust-metallicity values could be either related to superior SED forward-modeling, or simply be caused by incomplete coverage of the prior space due to the limited number of iterations.

Differences between the results from both codes can occur for a number of reasons. The statistical framework, prior choices, different models for contributions to the SED, as well as the choice of stellar spectral templates all influence the quality of the best-fit spectrum and its derived physical properties. While these do not seem to have a large effect on stellar masses (or even SFRs), the differences in the age-dust-metallicity space indicate that SED fitting results should not be over-interpreted in galaxy evolution studies, even with large baselines or quasi-spectroscopic resolutions. This is mainly due to the property degeneracies, but also caused by model and template choices in the applied SED code, as well as its statistical inference method.  

Overall, we find that while the PAUS narrowband photometry provides a large improvement on the accuracy of photometric redshifts, the effect on the estimation of physical properties of galaxy stellar populations is not very significant. With stellar masses and SFRs being mostly derived robustly with broadbands only, slight improvements can in general be expected for ages, metallicities, or dust-attenuation model parameters. Still, the choice of SED fitting code, stellar spectral library as well as for example dust or gas models have a large impact too. Thus, this approach requires caution in the interpretation of physical parameters from photometry in galaxy studies and shows the clear drawbacks of even well resolved photometry in comparison the spectroscopy.  

Finally, we have obtained a catalogue of new age, dust and metallicity measurements on the PAUS-COSMOS sample, which provides a more flexible and physically meaningful estimation, as it does not rely on simplified template fitting or fixing multiple SPS parameters (e.g., $Z_*$ or the dust attenuation curve) to reduce the dimensionality of the prior.

\begin{acknowledgements}
This work was funded in part by the Deutsche Forschungsgemeinschaft (DFG, German Research Foundation) under Germany’s Excellence Strategy – EXC-2094 – 390783311. \\

This work has been supported by the Polish National Agency for Academic Exchange (Bekker grant BPN/BEK/2021/1/00298/DEC/1), the European Union's Horizon 2020 Research and Innovation programme under the Maria Sklodowska-Curie grant agreement (No. 754510) \\

PR acknowledges the support by the Tsinghua Shui Mu Scholarship, the funding of the National Key R\&D Program of China (grant no. 2018YFA0404503), the National Science Foundation of China (grant no. 12073014 and 12350410365), the science research grants from the China Manned Space Project with No. CMS-CSST2021-A05, and the Tsinghua University Initiative Scientific Research Program (No. 20223080023). \\

GM acknowledges the support from the Collaborative Research Fund under Grant No. C6017-20G which is issued by the Research Grants Council of Hong Kong S.A.R. \\

CMB acknowledges support from STFC (ST/T000244/1,ST/X001075/1). \\

The PAU Survey is partially supported by MINECO under grants CSD2007-00060, AYA2015-71825, ESP2017-89838, PGC2018-094773, PGC2018-102021, PID2019-111317GB, SEV-2016-0588, SEV-2016-0597, MDM-2015-0509 and Juan de la Cierva fellowship and LACEGAL and EWC Marie Sklodowska-Curie grant No 734374 and no.776247 with ERDF funds from the EU Horizon 2020 Programme, some of which include ERDF funds from the European Union. IEEC and IFAE are partially funded by the CERCA and Beatriu de Pinos program of the Generalitat de Catalunya. Funding for PAUS has also been provided by Durham University (via the ERC StG DEGAS-259586), ETH Zurich, Leiden University (via ERC StG ADULT-279396 and Netherlands Organisation for Scientific Research (NWO) Vici grant 639.043.512), University College London and from the European Union's Horizon 2020 research and innovation programme under the grant agreement No 776247 EWC. The PAU data center is hosted by the Port d'Informaci\'o Cient\'ifica (PIC), maintained through a collaboration of CIEMAT and IFAE, with additional support from Universitat Aut\`onoma de Barcelona and ERDF. We acknowledge the PIC services department team for their support and fruitful discussions. \\

Based on observations obtained with MegaPrime/MegaCam, a joint project of CFHT and CEA/DAPNIA, at the Canada-France-Hawaii Telescope (CFHT) which is operated by the National Research Council (NRC) of Canada, the Institut National des Science de l'Univers of the Centre National de la Recherche Scientifique (CNRS) of France, and the University of Hawaii. This work is based in part on data products produced at TERAPIX and the Canadian Astronomy Data Centre as part of the Canada-France-Hawaii Telescope Legacy Survey, a collaborative project of NRC and CNRS. \\

This work has made use of CosmoHub. CosmoHub has
been developed by the Port d’Informació Científica (PIC), maintained through a collaboration of the Institut de Física d’Altes Energies (IFAE) and the Centro de Investigaciones Energéticas, Medioambientales y Tecnológicas (CIEMAT), and was partially funded by the ”Plan Estatal de Investigación Científica y Técnica y de Innovación” program of the Spanish government, has been supported by the call for grants for Scientific and Technical Equipment 2021 of the State Program for Knowledge Generation and Scientific and Technological Strengthening of the R+D+i System, financed by MCIN/AEI/ 10.13039/501100011033 and the EU NextGeneration/PRTR (Hadoop Cluster for the comprehensive management of massive scientific data, reference EQC2021- 007479-P) and by MICIIN with funding from European Union NextGenerationEU(PRTR-C17.I1) and by Generalitat de Catalunya \citep{Carretero2018, Tallada2020}. \\

PAUS data is hosted on CosmoHub from the v0.4 PAUS+COSMOS flux catalogue and the v0.4 PAUS+COSMOS photo-$z$ catalogue \citep{Eriksen2019, Tonello2019, Alarcon2021}. CFHTLS D2 field data from the T0007 final release \citep{Cuillandre2012} is publicly available from TERAPIX, the Canadian Astronomy Data Centre (CADC) and the Strasbourg astronomical Data Center (CDS).The COSMOS2020 catalogue with photometry, photo-$z$'s and physical properties can be accessed on \url{https://cosmos2020.calet.org} or on CosmoHub. The details and contents are described in \cite{Weaver2022}.
\end{acknowledgements}

\bibliographystyle{aa}
\bibliography{paperPAUS.bib}

\appendix
\onecolumn
\section{\label{apdx:posteriors}\prospector example posterior distributions}

\begin{figure}[h!]
    \centering
    \includegraphics[width=\columnwidth]{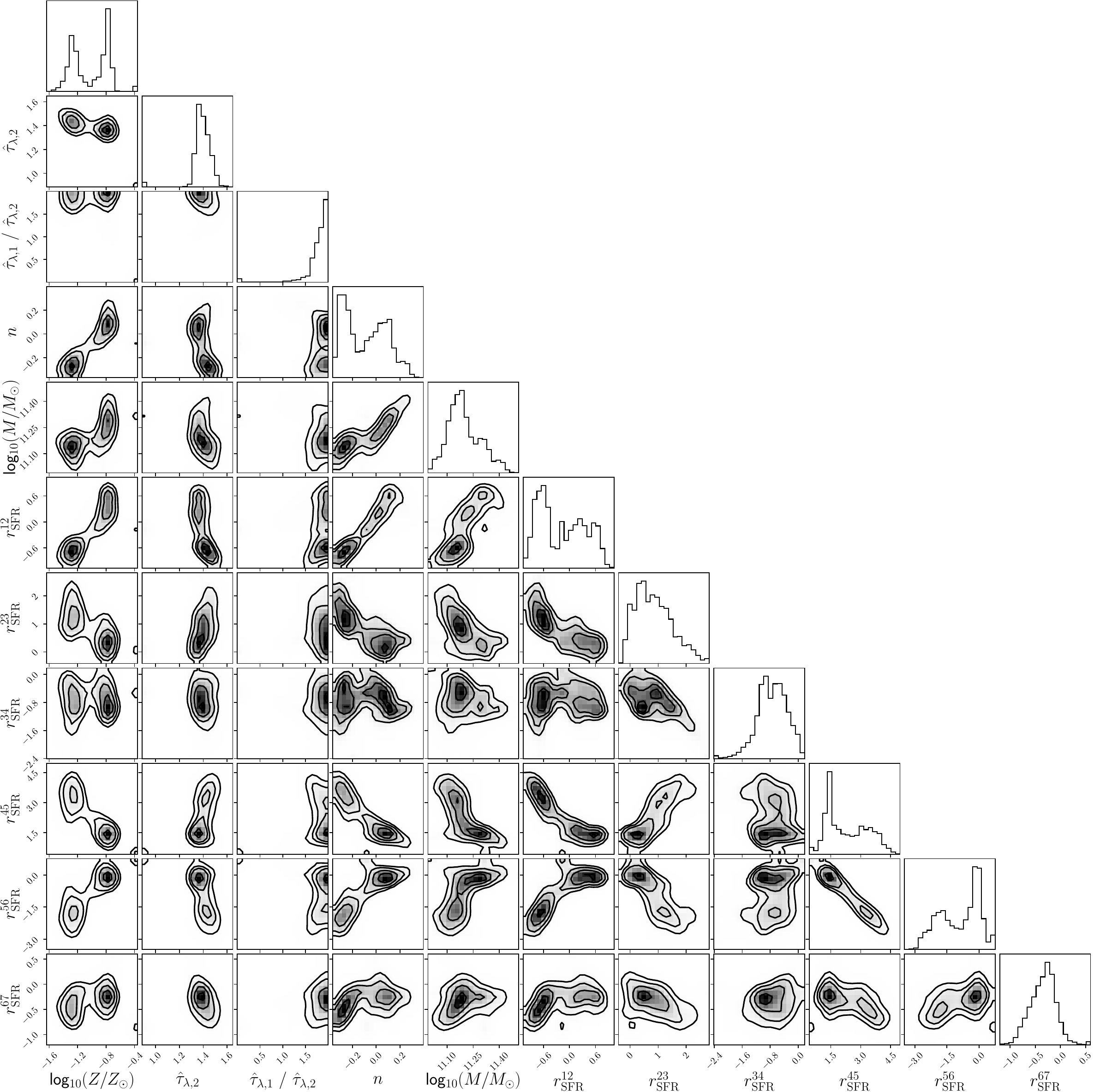}
    \caption{Posterior distribution of an example galaxy with non-converged properties from \prospector. The properties shown are the metallicity, the diffuse dust optical depths $\hat{\tau}_{\lambda}$, the ratio between diffuse dust and birth cloud optical depths, the dust slope modifier, $n$, the stellar mass, and the ratios of SFRs between adjacent age bins, which define the non-parametric SFH.}
    \label{fig:posterior-mult-peaks}
\end{figure}

Figure \ref{fig:posterior-mult-peaks} shows an example of a fit from \prospector where the chains do not converge, with many properties showing posteriors with multiple peaks. This is not due to short sampling, as the shown plot was obtained after MCMC with 64 walkers for 8192 iterations each, leading to an overall outcome of $>500\,000$ likelihood calls. The stellar mass is seemingly unaffected, with  a tight constraint, but metallicity, the slope modifier of the dust attenuation model and some SFR ratios clearly show a degenerate behavior. While this is not apparent for all galaxies, it still showcases how the correlation between dust, metallicity, and SFH can account for the large differences between simulated ground truth and fitted properties. Moreover, it proves that simply sampling longer does not necessarily improve the results as the degeneracy is not easily broken. 

\begin{figure}[h!]
    \centering
    \includegraphics[width=\columnwidth]{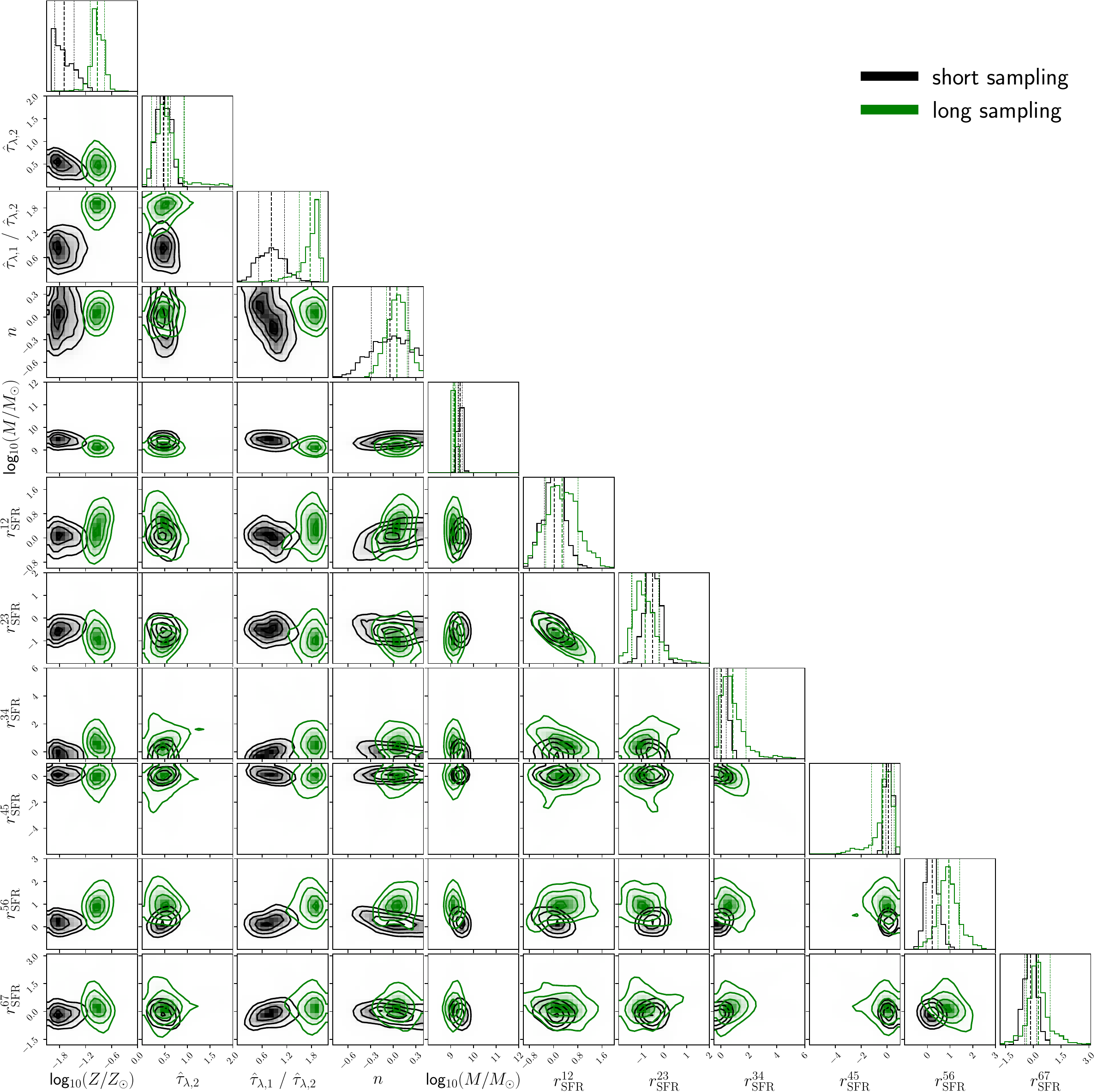}
    \caption{Comparison of the posterior distributions depending on the overall number of likelihood calls. The gray contour show the posteriors of the shown properties with 2048 iteration and 48 walkers, while the green contours depict the results from another run with iterations walkers and 64 walkers.}
    \label{fig:posterior-iteration-comp}
\end{figure}

\clearpage
Next, we  illustrate how an increase of likelihood calls can positively affect the (or change in general) the posterior distribution and thus the values of the best-fit SED model. Figure \ref{fig:posterior-iteration-comp} depicts the results from two \prospector runs on the same object, once with our nominal setup and once using 64 walkers with 8192 iteration each. While the stellar mass, as well as the SFR bin ratios of the non-parametric SFH do mostly agree within their $1\sigma$ contours, the ratio of diffuse and birth cloud dust optical depth and the stellar metallicity can shift significantly given more likelihood calls. This shows how the metallicity and dust property estimates can be unstable during the sampling process and show no clear convergence. However, this does not necessarily result in better results after longer sampling, due to the dust-age-metallicity degeneracy, where multiple points in this parameter space might have similar high likelihoods. Moreover, the shown example is not fully representative, as it is just a random sample. This can and will look significantly different, depending on the properties of the objects that is fitted. Still, it shows that we can have confidence in our stellar mass and SFR estimates, but caution is required for galaxy evolution studies in general when relying on SED fitting to infer specific physical parameters such as the mean stellar ages or metallicities. 

\clearpage

\section{\label{apdx:ssp-comp}Comparison of SED with varying SSPs}

\begin{figure}[h!]
    \centering
    \includegraphics[width=\columnwidth]{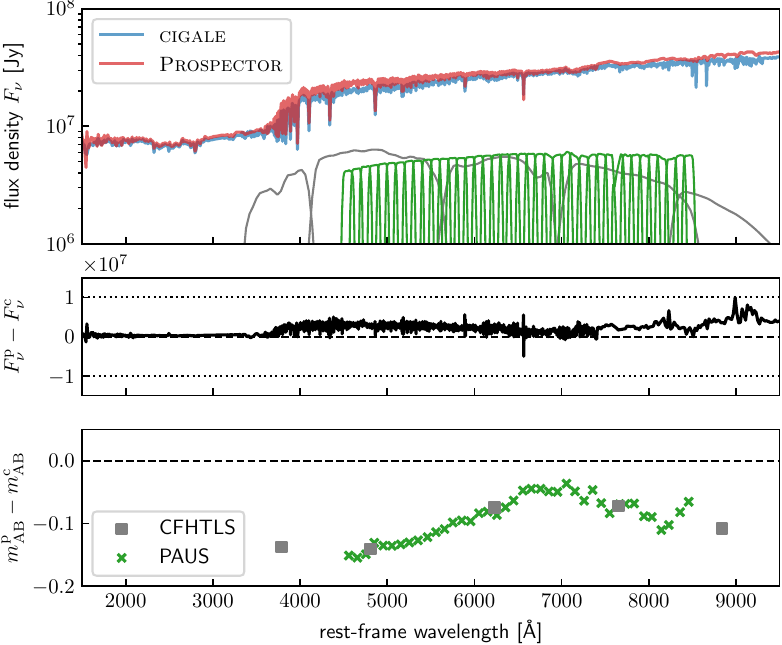}
    \caption{Comparison of rest-frame SED and magnitudes between BC03 (\cigale) and MILES (\prospector). The uppermost panel shows the rest-frame spectra, with the residual between the curves plotted below. The lower plot illustrates the magnitude differences in the CFHTLS and PAUS filter bands extracted from both SED simulations.}
    \label{fig:app-sed-comp}
\end{figure}

As aforementioned, the SSP templates available in \cigale and \prospector differ, leading to biases in the SED modeling given the same set of parameters. We here show in Fig. \ref{fig:app-sed-comp} a comparison plot between the output simulated composite stellar populations given the same set of input physical properties. The simulations assume a fixed metallicity $Z=Z_\odot$, a redshift $z=0.0$, and a delayed-$\tau$ shape of the SFH with $\langle t_*\rangle=5\,$Gyr, $\tau_*=2\,$Gyr, with no dust attenuation, dust emission or nebular emission. Thus, the models only differ by their applied stellar templates, with BC03 for \cigale and MILES for \prospector. The first subplot shows the overall rest-frame SEDs derived from both codes for the input parameter set, with the second subplot depicting the residual between the two curves. Below, we plot the magnitude differences for the observed photometry in the PAUS and CFHTLS bands extracted from the observed-frame spectra. Therein, we see that the MILES templates produce overall brighter magnitudes than BC03. The relative magnitude difference decreases with the wavelength up to the $g$-band, but increases again towards to near-infrared. Such differences in the UV, optical and infrared thus naturally lead to differences in the physical parameters, if the same photometry is fit using varying SSPs.

\end{document}